\documentclass[twocolumn,showpacs,preprintnumbers,amsmath,amssymb]{revtex4}

\usepackage{graphicx}   
\usepackage{bm}    

\newtheorem{Lemma}{Lemma}
\newtheorem{Theorem}{Theorem}

\newcommand{\Tr}{{\textrm{Tr}}}

\begin{document}

\title{Adiabatic Approximation for weakly open systems}

\author{Patrik Thunstr\"om}
\email{patrik.thunstrom@kvac.uu.se}
\author{Johan {\AA}berg}
\email{johan.aaberg@kvac.uu.se}
\author{Erik Sj\"oqvist}
\email{eriks@kvac.uu.se}
\affiliation{Department of Quantum Chemistry,
 Uppsala University, Box 518, SE-751 20 Uppsala, Sweden}

\date{\today}

\begin{abstract}
We generalize the adiabatic approximation to the case of open quantum
systems, in the joint limit of slow change and weak open system
disturbances. We show that the approximation is ``physically
reasonable'' as under wide conditions it leads to a completely
positive evolution, if the original master equation can be written on a
time-dependent Lindblad form. We demonstrate the approximation for a
non-Abelian holonomic implementation of the Hadamard gate, disturbed
by a decoherence process. We compare the resulting approximate
evolution with numerical simulations of the exact equation.
\end{abstract}


\maketitle
\section{Introduction}     

Recently, there has been a growing interest in the adiabatic
theorem \cite{Messiah} in the context of quantum information, in
particular for fault tolerant holonomic quantum computation
\cite{Zanardi}, and for the design of quantum adiabatic
algorithms \cite{FarhiGold,Farhi}. In this paper, we put forward a
type of adiabatic approximation with focus on how an ideal (unitary)
adiabatic evolution governed by a time-dependent Hamiltonian $H$ is
perturbed by open-system effects. We consider the effects of a
disturbance $D_{t/T}$ on a desired ideal evolution $\dot{\varrho} =
-i[H(t/T),\varrho]$, as a master equation $\dot{\varrho} =
L_{t/T}\varrho = -i[H(t/T),\varrho] + \Gamma D_{t/T} ( \varrho )$,
where $\Gamma$ gives the ``strength'' of the disturbance, and $T$ is
the run-time. In the ideal case $\Gamma=0$, the adiabatic
approximation decouples the evolution of the instantaneous eigenspaces
of $H$. In the present approximation, the eigenspace structure of
$H(t/T)$ still plays a role in that it determines what should decouple
in the adiabatic limit. It turns out that this in general limits the
applicability to systems that are weakly open, i.e., to small
$\Gamma$. The present study generalizes
Refs.~\cite{rompinto,pintorom}, as it allows degeneracy of the
Hamiltonian, which is an essential feature to obtain non-Abelian
holonomy effects in general \cite{Wilzee} and holonomic quantum
computation \cite{Zanardi} in particular.

The concept of adiabaticity in open systems has been addressed
recently by Sarandy and Lidar in Refs.~\cite{sarlidar,sali} for master
equations of the above type. In their approach, the adiabatic
approximation is characterized by a decoupling in terms of
instantaneous Jordan blocks of the superoperator $L_{t/T}$. In other
words, the decoupling is determined, in Ref.~\cite{sarlidar}, by the
total superoperator $L_{t/T}$, while this is determined by $H(t/T)$ in
the present approach.

The approach in Ref.~\cite{sarlidar} may be difficult to use in
certain applications. One example is the analysis of holonomic quantum
computation in the presence of open system effects. First of all it
should be noted that, although the Jordan decomposition always exists,
it could be challenging to determine it in practice for more than a
limited class of disturbances and systems. Furthermore, for the
approximation in Ref.~\cite{sarlidar} a new Jordan decomposition has
to be calculated for each choice of disturbance of the ideal gate.  In
the present approach, where the eigenspaces of the Hamiltonian are
primary, the approximate equation can be obtained irrespective of the
form of the disturbance $\Gamma D_{t/T}( \varrho )$. This is due to
the fact that the spectral decomposition of $H(t/T)$ is in general
known for holonomic implementations of quantum gates. For other
applications, however, the preferred method of approximation has to be
decided from the specific problem at hand.

Another generalization of adiabaticity to open systems has been
considered in \cite{Aab}, but for a specific type of open systems
in the context of quantum adiabatic search.

The structure of the paper is as follows. The approximation scheme 
is stated in the next section. In Sec.~\ref{sec:weak} we show
that the approximation can be obtained as an adiabatic weak
open-system limit of master equations. Section \ref{sec:compo} 
demonstrates that the approximation leads to a completely positive 
evolution under wide conditions. In Sec.~\ref{sec:appl} 
we apply the present approximation scheme to a decoherence model 
of a non-Abelian implementation of the Hadamard gate. Moreover, we 
compare with numerical solutions of the exact equation. The range of 
applicability of the approximation is discussed in Sec. \ref{sec:range}.  
The paper ends with the conclusions. 

\section{\label{sec:approx}The approximation}
We consider master equations of the following type 
\begin{equation}
\label{start}
\frac{d}{dt}\varrho(t) = 
-i[H(t/T),\varrho(t)] + \Gamma D_{t/T}\bm{(}\varrho(t)\bm{)},
\end{equation}
where $H(t/T)$ is a family of Hermitian operators, $D_{t/T}$ is a
superoperator, $T$ is the run-time of the evolution, and $\Gamma$ is
a strength parameter of the open system effect.  With the change of
variables $s = t/T$ one obtains
\begin{equation}
\label{basekv}
\frac{d}{ds}\rho(s) = 
-iT[H(s),\rho(s)] + \Gamma T D_{s}\bm{(}\rho(s)\bm{)},
\end{equation}
where $\rho(s) = \varrho(sT)$.  The superoperator $D_{s}$ is
assumed to be linear. In addition to purely technical
assumptions on $D_{s}$ such as sufficient smoothness with respect to
$s$, we assume that the solution $\varrho(t)$ does not grow without
bound with respect to some operator norm, as $t$ grows. This is 
necessary if $\varrho(t)$ is to be a density operator, and is achieved 
under suitable conditions if $D_{s}$ is on the Lindblad form.

We assume that the dimension of each eigenspace of $H(s)$ is 
fixed, so that we may write  
\begin{equation}
H(s) = \sum_{k=1}^{K} E_{k}(s)P_{k}(s).  
\end{equation}
Furthermore, for each $s$ we assume $E_{k}(s) \neq E_{l}(s)$ for all 
$k,l$ such that $k\neq l$, and $P_{k}(s)$ are projection operators such 
that $P_{k}(s)P_{l}(s) = \delta_{kl}P_{k}(s)$ and $\sum_{k}P_{k}(s) =
\hat{1}$.

Under conditions that are elucidated in Sec.~\ref{sec:weak}, the
adiabatic approximation of Eq.~(\ref{basekv}) takes the form
\begin{eqnarray}
\label{totalaapp}
\dot{\rho} & = &-i[TH(s)+Q(s),\rho] \\
& & +\Gamma T\sum_{klk'l'}g_{klk'l'}P_{k}(s)D_{s}\bm{(}P_{k'}(s)\rho
P_{l'}(s)\bm{)}P_{l}(s),\nonumber
\end{eqnarray}
where 
\begin{equation}
Q(s) = i\sum_{k}\dot{P}_{k}(s)P_{k}(s)
\end{equation}
is Hermitian (see Eq.~(\ref{sumPdotP})) and $g_{klk'l'}$ are $0$ or
$1$ depending on the pairwise eigenvalue differences
\begin{eqnarray}
\Delta_{kk'}(s) = E_{k}(s)-E_{k'}(s), 
\end{eqnarray}
as is described in Sec.~\ref{sec:offd}.  In the case of closed
evolution $\Gamma = 0$, we retain the standard adiabatic approximation
\cite{remark1}. 

An alternative form of Eq.~(\ref{totalaapp}) may be obtained by making
the change of variables
\begin{equation}
\label{change}
\widetilde{\rho}(s) = U(s)\rho(s)U^{\dagger}(s) ,
\end{equation}
where $U(s)$ is any sufficiently smooth family of unitary operators 
such that
\begin{equation}
\label{Pvillkor}
U(s)P_{k}(s)U^{\dagger}(s) = P_{k}(0), \quad \forall k . 
\end{equation}
In terms of $\widetilde{\rho}(s)$, Eq.~(\ref{basekv}) takes the 
form 
\begin{equation}
\label{huvudekv}
\dot{\widetilde{\rho}} = 
-iT[\widetilde{H}(s),\widetilde{\rho}(s)] 
-i[Z(s),\widetilde{\rho}(s)] 
+ \Gamma T \widetilde{D}_{s}(\widetilde{\rho}),
\end{equation}
where 
\begin{eqnarray}
\label{wHZdef}
\widetilde{H}(s) & = &  U(s)H(s)U^{\dagger}(s) =
 \sum_{k}E_{k}(s)P_{k}(0),\nonumber\\
Z(s) & = & i\dot{U}(s)U^{\dagger}(s),\nonumber\\
\widetilde{D}_{s}(\widetilde{\rho}) & = &  
U(s)D_{s}\bm{(}U^{\dagger}(s)\widetilde{\rho}(s)U(s)\bm{)}U^{\dagger}(s), 
\end{eqnarray}
and we have used that $U(s) \dot{U}^{\dagger}(s) = - \dot{U}(s) 
U^{\dagger}(s)$. We decompose the density operator as $\widetilde{\rho} =
\sum_{kl}\widetilde{\rho}^{(kl)}$, where $\widetilde{\rho}^{(kl)} =
P_{k}(0)\widetilde{\rho}P_{l}(0)$.   We refer to
$\widetilde{\rho}^{(ll)}$ as the ``diagonal'' terms, while we refer to
$\widetilde{\rho}^{(kl)}$, with $k\neq l$, as the ``off-diagonal''
terms.
The approximate Eq.~(\ref{totalaapp}) can be written as
\begin{eqnarray}
\label{nlan}
\frac{d}{ds}\widetilde{\rho}^{(kl)} & = & 
- iT\Delta_{kl}(s)\widetilde{\rho}^{(kl)}(s)
\nonumber\\
& &-iZ_{k}(s)\widetilde{\rho}^{(kl)}(s)+
i\widetilde{\rho}^{(kl)}(s)Z_{l}(s)
\nonumber\\ & & +\Gamma T
\sum_{k'l'}g_{klk'l'}P_{k}(0)
\widetilde{D}_{s}(\widetilde{\rho}^{(k'l')})P_{l}(0),
\end{eqnarray}
where $Z_{l}(s) = P_{l}(0)Z(s)P_{l}(0)$.  The properties of
$g_{klk'l'}$ imply that the diagonal terms $\widetilde{\rho}^{(ll)}$
always evolve according to the following equation
\begin{eqnarray}
\label{diagonal}
\frac{d}{ds}\widetilde{\rho}^{(ll)} & = & 
-i[Z_{l}(s),\widetilde{\rho}^{(ll)}(s)] 
\nonumber \\
& & + \Gamma T\sum_{k}P_{l}(0)
\widetilde{D}_{s}(\widetilde{\rho}^{(kk)})P_{l}(0).
\end{eqnarray}
The first term on the right-hand side of Eq.~(\ref{diagonal}) yields
the non-Abelian holonomy \cite{Wilzee} of the standard adiabatic approximation,
while the second term introduces a coupling between the diagonal terms
of the density operator. Equation (\ref{diagonal}) implies that for
the approximate evolution the diagonal terms always evolve
independently of the off-diagonal terms. In the simplest case where
$g_{klk'l'} = \delta_{kk'}\delta_{ll'}$, for $k\neq l$, the
off-diagonal terms evolve independently of each other and of the
diagonal terms.

We note that if $\widetilde{\rho}^{(kl)} (s)$ are the
solutions of Eq.~(\ref{nlan}), then
$U^{\dagger}(s)\widetilde{\rho}^{(kl)}(s)U(s) =
P_{k}(s)\rho(s)P_{l}(s)$, where $\rho(s)$ is the solution of
Eq.~(\ref{totalaapp}). This follows from the fact that
Eq.(\ref{nlan}) is equivalent to
Eq.~(\ref{totalaapp}), as is demonstrated in Sec.~\ref{sec:equiv}.

\section{\label{sec:weak}The approximation as an adiabatic 
weak open-system limit} Here, we put forward one possible way to
justify the above approximation scheme. First, we note that
Eq.~(\ref{huvudekv}) may be written as
\begin{equation}
\label{sftjf}
\frac{d}{ds}\widetilde{\rho}(s) = 
L^{(1)}_{s}\widetilde{\rho}(s) + L^{(2)}_{s}\widetilde{\rho}(s), 
\end{equation}
where
\begin{eqnarray}
\label{L1L2}
L_{s}^{(1)} & = & -iT[\widetilde{H}(s),\cdot] 
\nonumber,\\
L_{s}^{(2)} & = & -i[Z(s),\cdot] + \Gamma T \widetilde{D}_{s} . 
\end{eqnarray} 
Since $[\widetilde{H}(s),\widetilde{H}(s')]=0$, it follows that
$[L_{s}^{(1)},L_{s'}^{(1)}] = 0$. This implies that Eq.~(\ref{sftjf})
can be rewritten as the following integral equation
\begin{eqnarray}
\label{gene}
e^{\Lambda(s)}\overline{\rho}(s) & = &
e^{\Lambda(s)}\overline{\rho}(0) 
\nonumber \\ 
& & + e^{\Lambda(s)}\int_{0}^{s} e^{-\Lambda(s')}
L_{s'}^{(2)} \left( e^{\Lambda(s')}\overline{\rho}(s') \right)ds' , 
\nonumber\\
\Lambda(s) & = & \int_{0}^{s}L_{s}^{(1)}(s'')ds'' , 
\end{eqnarray}
where we have made the change of variables 
\begin{equation}
\label{basbyte}
\widetilde{\rho}(s) = e^{\Lambda(s)}\overline{\rho}(s) .  
\end{equation}
The superoperator $L_{s}^{(1)}$ is anti-Hermitian with respect to 
the Hilbert-Schmidt inner product $(A,B) = \Tr(A^{\dagger}B)$. 
Thus, $\exp[\Lambda(s)]$ is unitary, and we can 
rewrite Eq.~(\ref{gene}) as
\begin{eqnarray}
\label{erhn}
\overline{\rho}(s) & = & \overline{\rho}(0) 
\nonumber \\ 
& &+\int_{0}^{s}e^{-\Lambda(s')}L_{s'}^{(2)} \left( 
e^{\Lambda(s')} \overline{\rho}(s') \right)ds' . 
\end{eqnarray}
Note that $\sigma = \exp[\Lambda(s)]\sigma(0)$ is the solution 
of the equation $\dot{\sigma} = -iT[\widetilde{H}(s),\sigma]$. 
Since $\widetilde{H}(s)$ possesses a time-independent eigenbasis 
it follows that the corresponding evolution operator can be written 
as 
\begin{eqnarray}
V(s) & = & \sum_{k}\exp[-iTI_{k}(s)]P_{k}(0), 
\nonumber \\ 
I_{k}(s) & = & \int_{0}^{s}E_{k}(s')ds'.
\end{eqnarray}
Thus,
\begin{eqnarray}
e^{\Lambda(s)}\sigma & = & V(s)\sigma V^{\dagger}(s) 
\nonumber \\ 
 & = & 
\sum_{kl}e^{-iT\{I_{k}(s)-I_{l}(s)\}}P_{k}(0)\sigma P_{l}(0),
\end{eqnarray}
for every linear operator $\sigma$. We obtain
\begin{eqnarray}
\label{lautv}
\overline{\rho}(s) & = & e^{-\Lambda(s)}\widetilde{\rho}(s) 
\nonumber \\ 
 & = & \sum_{kl}e^{iT\{I_{k}(s)-I_{l}(s)\}} 
P_{k}(0) \widetilde{\rho}(s)P_{l}(0). 
\end{eqnarray}
If Eq.~(\ref{lautv}) is combined with Eq.~(\ref{erhn}) the result is
\begin{eqnarray}
\label{ekvationen}
\overline{\rho}(s) &= & \overline{\rho}(0)\\ &
 &+\sum_{klk'l'}\int_{0}^{s}e^{iTI_{klk'l'}(s')}\nonumber \\ & & \quad
 \times P_{k}(0)
 L_{s'}^{(2)}\bm{(}P_{k'}(0)\overline{\rho}(s')P_{l'}(0)\bm{)}
 P_{l}(0)ds', \nonumber
\end{eqnarray}
where
\begin{equation}
I_{klk'l'}(s) = I_{k}(s)-I_{l}(s) -I_{k'}(s)+ I_{l'}(s).
\end{equation}
Inserting Eq.~(\ref{L1L2}) into Eq.~(\ref{ekvationen}) yields 
\begin{eqnarray}
\label{ekvationensl}
\overline{\rho}(s) & = & \overline{\rho}(0)\\
& & -i\sum_{kk'}\int_{0}^{s}e^{iTI_{kk'}(s')}[P_{k}(0)
Z(s')P_{k'}(0),\overline{\rho}(s')]ds'\nonumber\\ & &+\Gamma
T\sum_{klk'l'}\int_{0}^{s}e^{iTI_{klk'l'}(s')}\nonumber\\ &
&\quad\times P_{k}(0)
D_{s'}\bm{(}P_{k'}(0)\overline{\rho}(s')P_{l'}(0)\bm{)}
P_{l}(0)ds',\nonumber
\end{eqnarray}
where we have introduced 
\begin{equation}
I_{kl}(s) = I_{k}(s)-I_{l}(s). 
\end{equation}

\subsection{\label{sec:diag}The diagonal terms}
The diagonal terms of Eq.~(\ref{ekvationensl}) read 
\begin{eqnarray}
\label{aoirae}
\overline{\rho}^{(ll)}(s) & = & 
P_l (0) \overline{\rho} (s) P_l (0) 
\nonumber \\ 
& = & \overline{\rho}^{(ll)}(0)
-i\int_{0}^{s}[Z_{l}(s'),\overline{\rho}^{(ll)}(s')]ds' 
\nonumber\\
& & + \Gamma T\sum_{k}\int_{0}^{s} P_{l}(0)
D_{s'}\bm{(}\overline{\rho}^{(kk)}(s')\bm{)} P_{l}(0)ds' 
\nonumber\\
& & + X_{d}(s) . 
\end{eqnarray}
Here, 
\begin{eqnarray}
\label{Fdef}
&&X_{d}(s)\nonumber\\ & &=\sum_{k:k\neq
l}\int_{0}^{s}e^{iTI_{lk}(s')}P_{l}(0)Z(s')
P_{k}(0)\overline{\rho}^{(kl)}(s')ds' 
\nonumber\\
& & -\sum_{k:k\neq l} \int_{0}^{s} e^{iTI_{kl}(s')}
\overline{\rho}^{(lk)} (s') P_{k}(0) Z(s') P_{l}(0) ds' \\ 
 & & +\Gamma T\!\!\!\!\sum_{k'l':k'\neq l'} \int_{0}^{s} 
e^{-iTI_{k'l'}(s')}P_{l}(0) 
D_{s'} \bm{(} \overline{\rho}^{(k'l')}(s') \bm{)} P_{l}(0) ds' , 
\nonumber
\end{eqnarray}
where we here have used that $I_{llk'l'}(s) = -I_{k'l'}(s)$.

We now show that the operator $X_{d}(s)$ vanishes in suitable 
limits of $T$ and $\Gamma$.  First, we cite Lemma 7.2.17 from 
\cite{complex}.

\begin{Lemma}
\label{lemmaett}
Suppose the function $h(s)$ is real valued, has a continuous second
derivative on the closed bounded interval $[0,1]$, and
$\frac{d}{ds}h(s)\neq 0$ for all $s\in[0,1]$. Let the function $f(s)$
have a continuous derivative on $[0,1]$. Then, for sufficiently large
$T$, there exists a constant $C$ such that
\begin{equation}
\label{asdfn}
\int_0^1 e^{iTh(s)}f(s)ds \leq CT^{-1}.
\end{equation}
\end{Lemma}

Since the integrands in Eq.~(\ref{Fdef}) all
 take the form $\exp[iTh(s)]F(s)$ it may be tempting to
use Lemma \ref{lemmaett}, or similar results like the Riemann-Lebesgue
Lemma \cite{RimLeb}, directly on these integrals. However, since $F$
depends on the solution $\overline{\rho}$, one should keep in mind
that the solution $\rho$ (and hence $\overline{\rho}$) depends on $T$,
and may contain fluctuations growing with $T$, which potentially may
cancel the averaging effect of the phase factors $\exp[iTh(s)]$. This
makes a direct use of Lemma \ref{lemmaett} dangerous when applied to
terms containing $\overline{\rho}$. In other words, we cannot allow
the function $f$ in Eq.~(\ref{asdfn}) to depend on $T$, neither
directly nor indirectly.  It is, however, quite straightforward to
avoid this problem in the present case.

Let $\{ |n\rangle \}_n$ be some fixed orthonormal basis,
independent of $s$, $T$, and $\Gamma$. With respect to this basis 
the first integral in Eq.~(\ref{Fdef}) can be written as 
\begin{equation}
\sum_{mn}|m\rangle\langle n|\int_{0}^{1}
e^{iTI_{lk}(s')}\bm{(}P_{l}(0)Z(s')P_{k}(0)\bm{)}_{mn} 
\overline{\rho}^{(kl)}_{nm}(s')ds' , 
\end{equation} 
which is a sum containing integrals on the form
\begin{equation}
\label{lkkbbms} 
\int_{0}^{s}e^{iTh(s')}f(s')K\bm{(}\overline{\rho}^{(kl)}(s')\bm{)}ds'. 
\end{equation}
Here, $h(s) = \pm[I_{k}(s)-I_{l}(s)]$ for some $k,l$ and $K$ denotes a
linear map from the operator to a matrix element in the matrix
representation of it, i.e., $K(\cdot) = \langle n|\cdot|m\rangle$ for
some $n,m$. Note that the function $f$ only depends on $s$, not on
$\overline{\rho}$ or $T$.  Similarly, the second term on the
right-hand side of Eq.~(\ref{Fdef}) can be written as a sum involving
integrals of the form (\ref{lkkbbms}).

Now, by partial integration of Eq.~(\ref{lkkbbms}) one obtains
\begin{eqnarray}
\label{hbszkj}
R_{Z}(s) & = &
K\bm{(}\overline{\rho}^{(kl)}(s)\bm{)}\int_{0}^{s}e^{iTh(s')}f(s')ds'\\
& & -
\int_{0}^{s}\!\!K\!\!\left(\frac{d}{ds}\overline{\rho}^{(kl)}(s')\right)
\int_{0}^{s'}\!e^{iTh(s'')}f(s'')ds''
ds'.\nonumber
\end{eqnarray}
By differentiation of Eq.~(\ref{ekvationensl}), and by use of the
standard operator norm $||\sigma|| = \sup_{||\psi||=1}||\sigma|\psi\rangle||$, one finds
\begin{eqnarray}
\left|\left|\frac{d}{ds}{\overline{\rho}}^{(kl)}(s')\right|\right| 
&\leq & A^{(d)}_{1} + B^{(d)}_{1}\Gamma T,
\end{eqnarray}
for some constants $A^{(d)}_{1}$ and $B^{(d)}_{1}$, where the index
$d$ signifies the diagonal terms. From Eq.~(\ref{hbszkj}) it follows
that
\begin{equation}
\label{kvvdk}
|R_{Z}(s)| \leq (1+ A^{(d)}_{1}+ B^{(d)}_{1}\Gamma
T)\bigg|\int_{0}^{s}e^{iTh(s')}f(s')ds'\bigg|,
\end{equation}
where we have used that $|K(\overline{\rho}^{(kl)})|\leq
||\overline{\rho}^{(kl)}||\leq 1$, as a consequence of the fact that
$\overline{\rho}(s)$ is a density operator. 

We note that $\frac{d}{ds} h(s) = E_{k}(s)-E_{l}(s)$ for all
$s\in[0,1]$, which is nonzero by assumption if $k\neq
l$. Furthermore, we assume that the family of Hermitian operators
$H(s)$ has an Hermitian continuous first derivative, which implies
that the eigenvalues $E_{k}(s)$ can be ordered, for each $s$, in such
a way that they have a continuous first derivative (see
Ref.~\cite{Rell}, pp.~44-45). Thus, the second derivative of $h(s)$ is
continuous if $H(s)$ has a continuous first derivative. Moreover, the
function $f$ does not depend on $T$, and has a continuous first
derivative if $Z(s)$ and $D_{s}$ has. We may thus apply Lemma
\ref{lemmaett} to the right-hand side of Eq.~(\ref{kvvdk}), from which
it follows that there exists some constant $C^{(d)}_{1}$ such that
\begin{equation}
|R_{Z}(s)|\leq C_{1}^{(d)}(1+ A_{1}^{(d)})T^{-1} +
C_{1}^{(d)}B_{1}^{(d)}\Gamma.
\end{equation}
The third integral in Eq.~(\ref{Fdef}) may be treated in the same way,
but including the extra factor $\Gamma T$, which results in terms
$R_{D}(s)$ bounded as
\begin{equation}
|R_{D}(s)|\leq C^{(d)}_{2}(1+ A^{(d)}_{2})\Gamma +
C_{2}^{(d)}B_{2}^{(d)}\Gamma^{2}T.
\end{equation} 
In total, we find that the norm (or, alternatively, the elements in some
matrix representation) of $X_{d}(s)$ is bounded as
\begin{equation}
||X_{d}(s)|| \leq A_{3}^{(d)}T^{-1} + B_{3}^{(d)}\Gamma +
C_{3}^{(d)}\Gamma^{2} T, 
\end{equation}
for some constants $A_{3}^{(d)}$, $B_{3}^{(d)}$, and $C_{3}^{(d)}$.

Next, we prove that the diagonal terms of the solution of the
exact Eq.~(\ref{ekvationensl}) converges to the solution of the
approximate equation of the diagonal terms, under certain conditions.

The set of operators $\sigma$ such that $\sum_{k}P_{k}(0)\sigma
P_{k}(0) = \sigma$, forms a linear subspace $\mathcal{L}$ of the space
of all linear operators on $\mathcal{H}$. Define
\begin{eqnarray*}
f_{d}(s,\sigma) & = & -i\sum_{l}P_{l}(0)[Z_{l}(s),P_{l}(0)\sigma
P_{l}(0)]P_{l}(0)\\ & & + \Gamma T\sum_{kl}
P_{l}(0)D_{s}\bm{(}P_{k}(0)\sigma P_{k}(0)\bm{)}P_{l}(0).
\end{eqnarray*}
For $\sigma,\sigma'\in\mathcal{L}$ and $s\in [0,1]$, we have  
\begin{eqnarray}
\label{dialip}
& &||f_{d}(s,\sigma)-f_{d}(s,\sigma')|| 
\nonumber \\ 
& &\leq \sum_{l}||P_{l}(0)[Z_{l}(s),P_{l}(0)(\sigma-\sigma')
P_{l}(0)]P_{l}(0)|| 
\nonumber \\
& & \quad + \Gamma T \sum_{kl}||P_{l}(0)D_{s} \! 
\bm{(}P_{k}(0)(\sigma-\sigma')P_{k}(0)\bm{)}P_{l}(0)|| 
\nonumber\\
& &\leq (F^{(d)} + G^{(d)}\Gamma T)||\sigma-\sigma'||,
\end{eqnarray}
for some constants $F^{(d)}$ and $G^{(d)}$. In the last inequality we
have used that $Z_{l}(s)$ and $D_{s}$ are continuous functions of
$s$ and that there exist maxima of $||Z_{l}(s)||$ and $|||D_{s}||| =
\sup_{||\sigma||=1}||D_{s}(\sigma)||$, the latter following from
 $[0,1]$ being a compact set. Note that the constants
$F^{(d)}$ and $G^{(d)}$ can be chosen independently of $\Gamma$ and
$T$. Equation (\ref{dialip}) means that $F^{(d)}+G^{(d)}\Gamma T$ is a
Lipschitz constant for $f_d$ on the set $[0,1]\times\mathcal{L}$.

Suppose that $\overline{\rho}^{a}_{d}(s)$ is the solution of the
approximate equation for the diagonal terms, i.e., Eq.~(\ref{aoirae})
with $X_{d}(s)\equiv 0$. Moreover, let
\begin{equation}
\overline{\rho}_{d}(s) = \sum_{l}P_{l}(0)\overline{\rho}(s)P_{l}(0)
 =\sum_{l}\overline{\rho}^{(ll)}(s),
\end{equation}
where $\overline{\rho}(s)$ is the exact solution of
Eq.~(\ref{ekvationensl}). We now intend to prove that
$||\overline{\rho}^{a}_{d}(s)-\overline{\rho}_{d}(s)||$ vanishes for
all $s$, in a suitable limit. 
The error $\mathcal{E}$, with respect to the standard operator norm, 
can be estimated as
\begin{eqnarray}
\mathcal{E}(s) & = & 
||\overline{\rho}^{a}_{d}(s)-\overline{\rho}_{d}(s)||\nonumber\\ & = &
\left|\left|\int_{0}^{s}\Big(f(s',\overline{\rho}^{a}_{d}(s'))-
f(s',\overline{\rho}_{d}(s'))\Big)ds' - X_{d}(s)\right|\right|\nonumber\\ &
\leq & ||X_{d}(s)|| +
\int_{0}^{s}||f(s',\overline{\rho}^{a}_{d}(s'))-
f(s',\overline{\rho}_{d}(s'))||ds'
\nonumber\\ & \leq & A_{3}^{(d)}T^{-1} + B_{3}^{(d)}\Gamma +
C_{3}^{(d)}\Gamma^{2} T\nonumber\\ & & + (F^{(d)} + G^{(d)}\Gamma T)
\int_{0}^{s}\mathcal{E}(s')ds'.
\end{eqnarray}
From the above inequalities one obtains an integral inequality for the
error $\mathcal{E}(s)$. This integral inequality can be shown
\cite{Amann} to have the solution
\begin{eqnarray}
\label{diagsc}
& &||\overline{\rho}^{a}_{d}(s)-\overline{\rho}_{d}(s)|| \\ & & \leq (
A_{3}^{(d)}T^{-1} + B_{3}^{(d)}\Gamma + C_{3}^{(d)}\Gamma^{2} T
)e^{s(F^{(d)} + G^{(d)}\Gamma T)}.\nonumber
\end{eqnarray} 
One can conclude that a sufficient condition for convergence of the
approximate and the exact solution is the simultaneous limits
$T\rightarrow \infty$ and $\Gamma\rightarrow 0$, under the condition
that $\Gamma T$ is bounded.

\subsection{\label{sec:offd}The off-diagonal terms}
The off-diagonal terms contain two types of phase factors, viz.,  
$\exp[iTI_{kl}(s)]$ and $\exp[iTI_{klk'l'}(s)]$. While $\frac{d}{ds} 
I_{kl}(s) = \Delta_{kl}(s)$ is always nonzero due to the assumption 
of distinct eigenvalues, the functions $\frac{d}{ds}I_{klk'l'}(s) =
\Delta_{kl}(s)-\Delta_{k'l'}(s)$ may be zero at isolated points, or
more systematically, even if $\Delta_{kl}(s)\neq 0$. 
Thus, the averaging effect leading to the adiabatic decoupling of the 
exact equation depends upon whether the graphs of the functions 
$\Delta_{kl}(s)$ avoid each other, cross, or coincide. We now study 
the following two physically reasonable special cases. 
\begin{itemize}
\item[(i)] For each pair $(k,l)$ and $(k',l')$ it holds 
that $\Delta_{kl}(s) = \Delta_{k'l'}(s),\quad \forall s\in[0,1]$, or
$\Delta_{kl}(s)\neq \Delta_{k'l'}(s),\quad \forall s\in[0,1]$.
\item[(ii)]  For each pair $(k,l)$ and $(k',l')$ it holds that
$\Delta_{kl}(s) = \Delta_{k'l'}(s),\quad \forall s\in[0,1]$, or
$\Delta_{kl}(s)\neq \Delta_{k'l'}(s)$ for all $s$, except possibly at
isolated points. At each such point $\widetilde{s}$ it
holds that $\frac{d}{ds} (\Delta_{kl}(\widetilde{s})- \Delta_{k'l'}
(\widetilde{s}))\neq 0$.
\end{itemize}
In the first case the condition says that the functions
$\Delta_{kl}(s)$ and $\Delta_{k'l'}(s)$ either coincide at all points,
or never cross.  In other words, the difference
$\Delta_{kl}(s)-\Delta_{k'l'}(s)$ is either zero or nonzero on the
whole interval $[0,1]$.  In the second case we allow the above
mentioned graphs to cross at isolated points. At those points where
they do cross we put a restriction on how they cross, in form of the
first derivative of $\Delta_{kl}-\Delta_{k'l'}$.

As the approximation depends on the behavior of the functions
$\Delta_{kl}$ we need to keep track of those which coincide
systematically. To do this in the above two special cases, we define 
\begin{equation}
\label{gdef}
g_{klk'l'} = 
\left\{ \begin{array}{lcr} 1 & \text{if} & \Delta_{kl}(s) = 
\Delta_{k'l'}(s),\quad\forall s\in[0,1], \\ 0 & \textrm{else}. &
\end{array} \right.
\end{equation}
Hence, $g_{klk'l'}=1$ only if the two functions $\Delta_{kl}(s)$ and
$\Delta_{k'l'}(s)$ coincide systematically.  Note that this definition
holds for all combinations of $k,l,k',l'$, including those involving
diagonal elements. It follows that 
\begin{equation}
\label{symm1}
g_{klk'l'}  =  g_{k'l'kl}, \quad g_{klk'l'}  =  g_{lkl'k'}, 
\end{equation}
and 
\begin{eqnarray}
\label{vilk1}
g_{klk'l}  =  \delta_{kk'}, & & 
g_{klkl'} = \delta_{ll'},
\nonumber\\
g_{kkk'l'}  =  \delta_{k'l'}, & &
g_{klk'k'} =  \delta_{kl} , 
\end{eqnarray}
where the latter conditions hold under the assumption 
$E_{k}(s)\neq E_{l}(s)$, $k\neq l$. 

The equation for the off diagonal term $\overline{\rho}^{(kl)}(s)$ can
be written
\begin{widetext}
\begin{eqnarray}
\label{totekv}
\overline{\rho}^{(kl)}(s) & = & \overline{\rho}^{(kl)}(0)
-i\int_{0}^{s}Z_{k}(s')\overline{\rho}^{(kl)}(s')ds'
+i\int_{0}^{s}\overline{\rho}^{(kl)}(s')Z_{l}(s')ds' 
\nonumber\\ & &+
\Gamma T \sum_{k'l'}g_{klk'l'}\int_{0}^{s}P_{k}(0)
D_{s'}\Big(\overline{\rho}^{(k'l')}(s')\Big) P_{l}(0)ds' + X_{o}(s),\\
\label{WTGdef}
X_{o}(s) & = & -i\sum_{k':k'\neq k}\int_{0}^{s}e^{iTI_{kk'}(s')}P_{k}(0)
Z(s')P_{k'}(0)\overline{\rho}^{(k'l)}(s')ds'\nonumber\\
& &  -i\sum_{k':k'\neq
l}\int_{0}^{s}e^{iTI_{k'l}(s')}\overline{\rho}^{(kk')}(s')P_{k'}(0)
Z(s')P_{l}(0)ds' 
\nonumber \\ 
 & & +\Gamma T\sum_{k'l':g_{klk'l'} =
0}\int_{0}^{s}e^{iTI_{klk'l'}(s')}P_{k}(0)
D_{s'}\Big(P_{k'}(0)\overline{\rho}(s')P_{l'}(0)\Big) P_{l}(0)ds'.
\end{eqnarray}
\end{widetext}
As in the proof of the approximate equation of the diagonal terms we
need a Lipschitz constant. Consider the linear operators
$\sigma$ which fulfills $\sum_{k,l:k\neq l}P_{k}(0)\sigma P_{l}(0) =
\sigma$. This operator subspace is denoted by $\mathcal{L}^{\perp}$. 
Define
\begin{eqnarray}
 & & f_{o}(s,\sigma) = 
\nonumber \\
 & & -i \sum_{kl:k\neq l} Z_{k}(s') P_{k}(0) \sigma P_{l}(0) ds' 
\nonumber \\ 
 & & +i\sum_{kl:k\neq l} P_{k}(0) \sigma P_{l}(0) Z_{l}(s') \\ 
 & & + \Gamma T \sum_{kl:k\neq l}\sum_{k'l'}g_{klk'l'}P_{k}(0) 
D_{s'}\bm{(}P_{k'}(0)\sigma P_{l'}(0)\bm{)} P_{l}(0). 
\nonumber
\end{eqnarray}
With a similar reasoning as for the diagonal terms we obtain
\begin{equation}
\label{lipso}
||f_{o}(s,\sigma)-f_{o}(s,\sigma')|| \leq (F^{(o)}+G^{(o)}\Gamma
T)||\sigma-\sigma'||,
\end{equation}
for all $s\in[0,1]$ and $\sigma,\sigma'\in\mathcal{L}^{\perp}$. 
Here, $F^{(o)}$ and $G^{(o)}$ are constants. We further define 
\begin{eqnarray}
\overline{\rho}_{o}(s) = 
\sum_{kl:k\neq l}P_{k}(0)\overline{\rho}(s) P_{l}(0)
 =\sum_{kl:k\neq l}\overline{\rho}^{(kl)}(s), 
\end{eqnarray}
where $\overline{\rho}(s)$ is the solution of
Eq.~(\ref{ekvationensl}).  Similarly we let
\begin{eqnarray}
\overline{\rho}_{o}^{a}(s) = 
\sum_{kl:k\neq l} \overline{\rho}^{(kl)}_{a}(s), 
\end{eqnarray} 
where $\overline{\rho}^{(kl)}_{a}(s)$ are the solutions of
Eq.~(\ref{totekv}), with $X_{o}(s) \equiv 0$. Clearly, both
$\overline{\rho}_{o}(s)$ and $\overline{\rho}_{o}^{a}(s)$ belong to
$\mathcal{L}^{\perp}$.

\subsubsection{Case (i)}
Here, the functions $\Delta_{kl}(s)$ and $\Delta_{k'l'}(s)$ 
either coincide at all points or never cross. Following the 
reasoning of the diagonal case, one finds that
\begin{eqnarray}
||X_{o}(s)|| \leq A^{(o1)}_{3}T^{-1} + B^{(o1)}_{3} \Gamma +
C^{(o1)}_{3} \Gamma^{2} T 
\end{eqnarray}
and we may use the Lipschitz condition in Eq.~(\ref{lipso}) to obtain 
\begin{eqnarray} & &
||\overline{\rho}_{o}(s)-\overline{\rho}_{o}^{a}(s)||\\ & & \leq
(A^{(o1)}_{3}T^{-1} + B^{(o1)}_{3} \Gamma + C^{(o1)}_{3} \Gamma^{2}
T)e^{s(F^{(o)}+ G^{(o)}\Gamma T)},\nonumber
\end{eqnarray}
for all $s\in[0,1]$. Thus, as for the diagonal terms, we find the  
conditions $T\rightarrow \infty$, $\Gamma\rightarrow 0$, and
$\Gamma T$ bounded, for convergence of the approximate and  
exact solution. 

\subsubsection{Case (ii)}
In this case we allow the graphs of the functions $\Delta_{kl}(s)$ to
cross, but only at isolated points, and with a nonzero angle. 
Consider two distinct pairs $(k,l)$ and $(k',l')$. It follows
that the function $\Delta_{kl}(s) - \Delta_{k'l'}(s)$ has only
isolated zeros. Since the interval $[0,1]$ is compact and since 
$\Delta_{kl}(s) - \Delta_{k'l'}(s)$ is continuous, there can only 
be a finite number of isolated zeros. We may
partition the interval $[0,1]$ into subintervals were each subinterval
has at most one zero of $\Delta_{kl}(s) - \Delta_{k'l'}(s)$ in its
interior.  Due to the zeros of $\Delta_{kl}(s) - \Delta_{k'l'}(s)$,
Lemma \ref{lemmaett} is no longer applicable, but we may instead use
the stationary phase theorem, which we cite from \cite{complex}
(Theorem 7.2.10). Note that we here present a weakened form of the
theorem, which precisely covers the aspects we need.

\begin{Theorem} 
\label{stationary}
Let $h(s)$ be analytic in a neighborhood of the closed bounded
interval $[a,b]$ and be real on $[a,b]$. Let $f(s)$ have a continuous
first derivative on $[a,b]$. If $\frac{d}{ds} h(s) = 0$ at exactly one
point $s_{0}\in (a,b)$ and if the second derivative of $h$ at $s_{0}$
is nonzero, then for sufficiently large $T$ there exists a constant
$D$ such that
\begin{equation}
\label{stat}
\int_a^b e^{iTh(s)}f(s)ds\leq DT^{-1/2}.
\end{equation}
\end{Theorem}

We write $X_{o}(s)$ as a sum of integrals of the form
(\ref{lkkbbms}), each of which is decomposed into integrals on
subintervals. In the present case, we may reason in the same way as in
the steps from Eq.~(\ref{hbszkj}) to Eq.~(\ref{kvvdk}), with the
exception that each integral spans only a subinterval. Note that we
only need to use Theorem \ref{stationary} on neighborhoods of the
points where the functions $\Delta_{kl}$ cross. On the rest of the
interval we may use Lemma \ref{lemmaett}.  Thus, in order to use
Theorem \ref{stationary}, the eigenvalues $E_{k}(s)$ only have to 
be analytic functions of $s$ in a neighborhood
of each point $s_{0}$ where $\Delta_{kl}(s_{0})-\Delta_{k'l'}(s_{0}) =
0$.  Since the Hilbert space is finite-dimensional and since $H(s)$ is
Hermitian, this is the case if $H(s)$ is analytic in a neighborhood of
each $s_{0}$ (see Ref.~\cite{Rell}, pp.~33-34, or Ref.~\cite{reed}, Theorem 
XII.3). We also require that $Z(s)$ and $D_{s}$ have continuous 
first derivatives in $s$.

The value of the integral
\begin{equation}
\left|\int_{a}^{b}e^{iTh(s)}f(s)ds\right|
\end{equation}
 is $O(T^{-1})$ if the subinterval $[a,b]$ does not contain a zero
of $\Delta_{kl}(s) -
\Delta_{k'l'}(s)$, and $O(T^{-1/2})$ if it does. When summing up the
contributions from the subintervals, it follows that the value of the
integral in Eq.~(\ref{kvvdk}), for sufficiently large $T$ can be
bounded as $D^{(o2)}_{1}T^{-1/2}$, for some constant $D^{(o2)}_{1}$.
Thus, the first two terms on the right-hand side of Eq.~(\ref{WTGdef}) are
bounded by a finite sum of expressions on the form
\begin{equation}
|R_{Z}(s)|\leq (1+A^{(o2)}_{1} + B^{(o2)}_{1}\Gamma
T)D^{(o2)}_{1}T^{-1/2},
\end{equation}
for sufficiently large $T$. 
Similarly, for the third term on the right-hand side of
Eq.~(\ref{WTGdef}) gives a finite sum of bounds of the form
\begin{equation}
|R_{D}(s)|\leq (1+A^{(o2)}_{2} + B^{(o2)}_{2}\Gamma
T)D^{(o2)}_{2} \Gamma T^{1/2}.
\end{equation}
Thus, for sufficiently large $T$ we obtain
\begin{displaymath}
||X_{o}(s)|| \leq A^{(o2)}_{3}T^{-1/2} + B^{(o2)}_{3}\Gamma T^{1/2} +
C^{(o2)}_{3}\Gamma^{2}T^{3/2}.
\end{displaymath}
By combining this with the Lipschitz condition (\ref{lipso}), one obtains 
\begin{eqnarray}
\label{o2sca}
& ||\overline{\rho}_{o}(s)-\overline{\rho}_{o}^{a}(s)||&\leq
(A^{(o2)}_{3}T^{-1/2} + B^{(o2)}_{3}\Gamma T^{1/2} \nonumber\\ & & +
C^{(o2)}_{3}\Gamma^{2}T^{3/2} ) e^{s(F^{(o)}+ G^{(o)}\Gamma T)},
\end{eqnarray}
for all $s\in[0,1]$.
Since, $\Gamma T^{1/2} = (\Gamma T)T^{-1/2}$ and $\Gamma^{2}T^{3/2} =
(\Gamma T)^{2}T^{-1/2}$, it is sufficient with the simultaneous
conditions $T\rightarrow
\infty$, $\Gamma \rightarrow 0$, and $\Gamma T$ bounded,
for the error to vanish. Although we obtain the same conditions as in
case (i), Eq.~(\ref{o2sca}) nevertheless indicates worse scaling
properties of the error than Eq.~(\ref{diagsc}) does. This point will
be discussed further in Sec.~\ref{sec:scales}.
   
\subsection{\label{sec:equiv}The approximate equations}
In Secs.~\ref{sec:diag} and \ref{sec:offd} we have motivated the
approximate equations for diagonal as well as off-diagonal terms
$\overline{\rho}^{(kl)}(s)$. For the diagonal terms the approximate
equation is Eq.~(\ref{aoirae}) with $X_{d}(s)\equiv 0$. For the
off-diagonal terms the approximate equation is Eq.~(\ref{totekv}) with
$X_{o}(s)\equiv 0$. One may transform the integral equations into
differential equations, followed by a change of variables back to
$\widetilde{\rho}^{(kl)}(s)$. With use of
the definition of $g_{klk'l'}$, this results in
Eqs.~(\ref{diagonal}) and (\ref{nlan}), for the diagonal and the
off-diagonal terms, respectively. Note that Eq.~(\ref{nlan}) holds,
 not only for the
off-diagonal terms, but for the diagonal terms as well. This is the
case since the expression in Eq.~(\ref{nlan}) reduces to
Eq.~(\ref{diagonal}), due to Eq.~(\ref{vilk1}), if we consider the
diagonal terms.

The transformation to Eq.~(\ref{totalaapp}) from Eq.~(\ref{nlan}) is
straightforward for the dissipator. The only part which may need
comment is the operator $Q(s)$ in Eq.~(\ref{totalaapp}). If one
transforms from the variable $\widetilde{\rho}(s)$, back to the
variable $\rho(s)$, combining all the terms, one obtains
\begin{eqnarray}
\label{intermediate}
\dot{\rho} & = & -iT[H(s),\rho] 
\nonumber\\ 
& & + \Gamma T\sum_{klk'l'}g_{klk'l'}P_{k}(s)D_{s}\bm{(}P_{k'}(s)\rho
P_{l'}(s)\bm{)}P_{l}(s) 
\nonumber \\ 
& & - \sum_{k}P_{k}(s)\dot{P}_{k}(s)\rho -
\rho\sum_{k} \dot{P}_{k}(s) P_{k}(s).
\end{eqnarray} 
By differentiating $P_{k}^{2}(s) = P_{k}(s)$ one obtains
$\dot{P}_{k}(s)P_{k}(s) + P_{k}(s)\dot{P}_{k}(s) = \dot{P}_{k}(s)$. If
this expression is summed over $k$ and is combined with the fact that
$\sum_{k}\dot{P}_{k}(s) = 0$, the result is
\begin{equation}
\label{sumPdotP}
\sum_{k} P_{k} (s) \dot{P}_{k}(s) = -\sum_{k} \dot{P}_{k}(s)P_{k}(s). 
\end{equation}
By combining this expression with Eq.~(\ref{intermediate}) one obtains
Eq.~(\ref{totalaapp}). 

\subsection{\label{sec:scales}Time scales}
For the diagonal terms, as well as for the off-diagonal terms in case
(i), we have found that the error between the solution of the exact
equation and the solution of the approximate equation satisfies a
bound of the form
\begin{equation}
\label{cond}
\mathcal{E}\leq (AT^{-1}+ B\Gamma + C\Gamma^{2}T)e^{F+G\Gamma T}.
\end{equation}
In view of the limiting processes considered in the previous sections,
a physical interpretation of this condition might be to assume that
the strength parameter $\Gamma$ depends on the run-time $T$. If
$\Gamma = \alpha/T$, with $\alpha\geq 0$ a constant independent of
$T$, then the error would go to zero when $T\rightarrow\infty$.  This,
however, paints our abilities to control open-system effects in a bit
too rosy colors. In practice, the open-system effects are often
residual uncontrollable errors and the strength $\Gamma$ is given by
the situation at hand, and we have no possibility to decrease $\Gamma$
as $T$ increases.

On the other hand, from Eq.~(\ref{cond}) it is quite clear that the
approximation is good if the run time $T$ is sufficiently large, the
characteristic time scale of the open-system effects $\Gamma^{-1}$ is
sufficiently large, and the run time $T$ is in the order of or smaller
than $\Gamma^{-1}$.  Unlike the standard adiabatic approximation where
the error can be made arbitrarily small by increasing the run-time,
the present approximation appears to be limited, since for a given
open-system strength $\Gamma$, the error cannot be made arbitrarily
small as the run-time has to be at the same order or smaller than the
characteristic time scale of the open-system effects
\cite{remark2}. One should keep in mind, though, that we only have
obtained sufficient conditions, not necessary conditions, for the
accuracy of the approximation. As pointed out in Sec.\ref{sec:range},
these sufficient conditions may in some cases be unnecessarily
pessimistic.

For the off-diagonal terms in case (ii), we similarly obtained the
condition $T\rightarrow\infty$, $\Gamma\rightarrow 0$, and $\Gamma T$
bounded. However, this does not tell us at what rate the error
decreases. One may compare Eqs.~(\ref{o2sca}) and (\ref{cond}). As an
example, consider those terms that solely depend on $T$, and not
$\Gamma$. This part scales like $T^{-1/2}$ and $T^{-1}$ in
Eqs.~(\ref{o2sca}) and Eq.~(\ref{cond}), respectively. Thus, while
both these parts goes to zero when the run-time $T$ increases, the
rate is slower for case (ii) than for the diagonal terms and case
(i). Similarly one may compare the other terms, in Eqs.~(\ref{o2sca})
and (\ref{cond}), containing combinations of $T$ and $\Gamma$. Again
one finds that the scaling of the error with increasing $T$ and
decreasing $\Gamma$ is worse in case (ii) than for the diagonal terms
and case (i). This suggests that the range of applicability of the
approximation is tighter in case (ii).

One may note that the constants $A$, $B$, $C$, $G$, and $F$ in
Eq.~(\ref{cond}) play an important role as they ``set the scales'', in
the sense that they determine what ``large $T$'' and ``small $\Gamma$''
means. We have avoided to give explicit estimates of these
constants. It would be possible to perform the derivations in the
previous sections in such a way that estimates of these constants are
obtained. However, it seems a better strategy to derive such constants
more specifically for the system and the initial conditions at
hand. 

In essence, we have shown that there exists a region of large $T$
and small $\Gamma$ where the approximation is good, but we have not
determined how large $T$ and how small $\Gamma$ must be. This is
analogous to perturbation theory where one knows the approximation to
be good if the perturbation parameter is sufficiently small, but
usually one does not know how small ``sufficiently small'' is.

\section{\label{sec:compo}Complete positivity}
So far we have assumed very little about the exact nature of the
disturbance $D_{s}$. Except that $D_{s}$ should be linear as a
superoperator and be sufficiently smooth as a function of $s$, we
have only required that it should lead to an evolution which keeps the
solution $\rho(s)$ bounded.  In this section we investigate in more
detail what evolution the approximation gives rise to, and we do so
for a restricted class of superoperators $D_{s}$.

For an important class of master equations $\dot{\varrho} = L\varrho$,
the superoperator $L$ can be written on the (time-independent) Lindblad form
\cite{Lindblad}. If $L$ is bounded, then the Lindblad form guarantees
that the resulting evolution is trace preserving and completely
positive \cite{Lindblad,Vitt}. To be more precise, the master equation
induces a one-parameter family of linear maps $\Lambda_{x}$ such that
$\rho(s_2) = \Lambda_{s_2-s_1}\rho(s_1)$, for $s_2\geq s_1$. Each
$\Lambda_{x}$ is trace preserving and completely positive if $L$ can
be written on the Lindblad form. The complete positivity guarantees
that the evolution maps density operators to density operators, even
if the evolution acts on one member of an entangled pair of systems
\cite{Kraus}.

If the superoperator $L$ is time-dependent we instead obtain a
two-parameter family of linear maps $\Lambda_{s_2,s_1}$ such that
$\rho(s_2) = \Lambda_{s_2,s_1}\rho(s_1)$, for $s_2\geq s_1$. 
In the finite-dimensional case it can be shown that a sufficient
condition for $\Lambda_{s_2,s_1}$ to be completely positive is that
the time-dependent superoperator $L_{s}$ can be written on a
time-dependent Lindblad form, and that $L_{s}$ has a continuous first
derivative on the interval $[0,1]$.
For discussions on the complete positivity of the
dynamical maps generated by time-dependent Lindbladians, see
Refs.~\cite{Alle,Lendi}.  Note that the Lindblad form of the master
equation is not necessary in order to obtain complete positivity, or
 positivity of the dynamical maps. More general
master equations which are non-local in time (integro-differential
equations) have been considered in the literature (see, e.g.,
\cite{Wilkie}). Moreover, time-local equations $\dot{\rho}(s) =
L_{s}\rho(s)$, where $L_{s}$ is not of the lindblad form have also
been considered (see, e.g., \cite{Breuer}).
In this section we assume that the disturbance $D_{s}$ can be written
on the time-dependent Lindblad form. It may be possible to generalize
the reasoning in this section to the type of master equations
considered in
\cite{Breuer}. This question is, however, not treated here.

Here it is shown that if the superoperator $D_{s}$ can be written on
the time-dependent Lindblad form, then the approximate equation can
also be written on the time-dependent Lindblad form. Thus, under
suitable conditions it follows that the approximate evolution is
``physically reasonable'' in the sense that it is trace preserving and
completely positive. 

Suppose $D_{s}(\rho)$ can be written on the time-dependent Lindblad form
\begin{eqnarray}
\label{Ds}
D_{s}(\rho) &=& -i[F(s),\rho] + \sum_{n}V_{n}(s)\rho
V_{n}^{\dagger}(s)\\ & &-\frac{1}{2}\sum_{n}
V_{n}^{\dagger}(s)V_{n}(s)\rho -\frac{1}{2}\rho\sum_{n}
V_{n}^{\dagger}(s)V_{n}(s),\nonumber
\end{eqnarray}
where $F(s)$ is Hermitian. To calculate the term of
Eq.~(\ref{totalaapp}) involving $D_s$, we use Eq.~(\ref{Ds}) together
with the conditions in Eq.~(\ref{vilk1}), to obtain the approximate
equation $\dot{\rho} = -i[TH(s)+ Q(s),\rho] + \Gamma
TR_{s}(\rho)$, where
\begin{eqnarray}
\label{njknd}
R_{s}(\rho) & = & 
\sum_{klk'l'}g_{klk'l'} P_{k}(s)D_{s}\bm{(} P_{k'}(s) \rho
P_{l'}(s) \bm{)} P_{l}(s) 
\nonumber \\ 
 & = & -i\sum_{k} \big[ P_{k}(s)
F(s)P_{k}(s), \rho \big] 
\nonumber \\ 
 & & + \sum_{klk'l'n}g_{klk'l'}P_{k}(s)V_{n}(s)P_{k'}(s)\rho P_{l'}(s)
V_{n}^{\dagger}(s)P_{l}(s) 
\nonumber \\ 
 & & -\frac{1}{2} \sum_{kn} P_{k}(s) V_{n}^{\dagger}(s) 
V_{n}(s) P_{k}(s) \rho
\nonumber \\ 
 & & -\frac{1}{2}\sum_{ln}\rho P_{l}(s)V_{n}^{\dagger}(s) 
V_{n}(s)P_{l}(s).
\end{eqnarray}
One may note the following 
\begin{eqnarray}
\label{ghsns}
 & & \sum_{k}P_{k}(s)V_{n}^{\dagger}(s)V_{n}(s)P_{k}(s) \\ 
 & & =\sum_{klk'l'}g_{klk'l'}P_{l'}(s)V_{n}^{\dagger}(s)
P_{l}(s)P_{k}(s)V_{n}(s)P_{k'}(s) 
\nonumber,
\end{eqnarray}
which follows from Eq.~(\ref{vilk1}).
By combining Eqs.~(\ref{njknd}) and (\ref{ghsns}) the result is
\begin{eqnarray}
\label{dfddfb}
 & & R_{s}(\rho) = 
-i\sum_{k} \big[ P_{k}(s) F(s)P_{k}(s),\rho \big] \\ 
 & & +
\sum_{klk'l'n}g_{klk'l'}P_{k}(s)V_{n}(s)P_{k'}(s)\rho P_{l'}(s)
V_{n}^{\dagger}(s)P_{l}(s) 
\nonumber \\ 
 & & -\frac{1}{2}
\sum_{klk'l'n}g_{klk'l'}P_{l'}(s)V_{n}^{\dagger}(s)
P_{l}(s)P_{k}(s)V_{n}(s)P_{k'}(s)\rho 
\nonumber\\
 & & -\frac{1}{2}\sum_{klk'l'n}g_{klk'l'}\rho
P_{l'}(s)V_{n}^{\dagger}(s)P_{l}(s)P_{k}(s)V_{n}(s)P_{k'}(s). 
\nonumber
\end{eqnarray}
Define a matrix $G$ with elements $G_{kk',ll'} = g_{klk'l'}$.  
$G$ is symmetric due to Eq.~(\ref{symm1}), which implies that 
$G$ is diagonalizable such that $G_{kk',ll'} = \sum_{m} \lambda_{m} 
c_{kk'}^{(m)} c_{ll'}^{(m)\ast}$. This can be used to show that 
Eq.~(\ref{dfddfb}) can be rewritten as
\begin{eqnarray}
R_{s}(\rho) & = & 
-i\sum_{k} \big[ P_{k}(s) F(s)P_{k}(s),\rho \big] 
\nonumber \\
 & & + \sum_{n}\sum_{m}M_{n}^{(m)}(s)\rho M_{n}^{(m)\dagger}(s) 
\nonumber \\ 
 & & -\frac{1}{2}\sum_{n}\sum_{m}
M_{n}^{(m)\dagger}(s)M_{n}^{(m)}(s)\rho 
\nonumber \\ 
 & & -\frac{1}{2}\sum_{n}\sum_{m}\rho M_{n}^{(m)\dagger}(s)M_{n}^{(m)}(s),
\end{eqnarray}
where
\begin{equation}
M_{n}^{(m)}(s) =
\sum_{kk'}\sqrt{\lambda_{m}}c_{kk'}^{(m)}P_{k}(s)V_{n}(s)P_{k'}(s).
\end{equation}
Hence, we have shown that the approximate equation can be written on
the time-dependent Lindblad form.

\section{\label{sec:appl}Example: Non-Abelian Holonomy} 
Holonomic quantum computation \cite{Zanardi} is a recently proposed
approach to quantum circuits using the idea of adiabatic
evolution. Here, we wish to apply the present approximation scheme 
for weak open-system effects in holonomic single-qubit rotation gates.

We consider a four-level system consisting of three ground states 
$0$, $1$, and $a$ whose coupling to an excited state $e$ is 
modeled by the Hamiltonian \cite{Unan}
\begin{equation}
\label{Hamiltonian}
H(s) =|e\rangle \big( \langle 0| \omega_0(s) + 
\langle 1| \omega_1(s) +
\langle a| \omega_a(s) \big) + \mathrm{H.c} .  
\end{equation}
Here, $s = t/T$, with $T$ being the run-time of the process, and
$\omega_{0}$, $\omega_{1}$, and $\omega_{a}$ are tunable, possibly
complex-valued, coupling parameters. For each $s$, $H$ possesses a
doubly degenerate zero-energy (dark) eigensubspace spanned by $|\chi_1
\rangle$ and $|\chi_2 \rangle$ and two bright eigenvectors 
$|\chi_3 \rangle$ and $|\chi_4 \rangle$, the latter with energies 
$\pm \omega$, where
\begin{equation}
\label{smomedef}
\omega = \sqrt{|\omega_0|^2+|\omega_1|^2+|\omega_a|^2}.
\end{equation}
 This type of system
is found in various implementations of holonomic gates, including
ion-traps \cite{Duan}, Josephson junctions \cite{Faoro},
semiconductor quantum dots \cite{solzanzanros}, and neutral atoms in
cavities \cite{RecCal}.

In the present investigation $\omega(s)$ is chosen to be constant. It
follows that we may measure energy in units of $\omega$, and thus let
the vector $[\omega_{0}(s),\omega_{1}(s),\omega_{a}(s)]$ to be of unit
length.  Since $\hbar = 1$ it follows that we measure
time, and especially the run-time, in units of $\omega^{-1}$. We use
this convention in the rest of this section.

Holonomic single-qubit
rotations acting on the computational space spanned by $|0\rangle$ and
$|1\rangle$ may be obtained in adiabatic transport of the doubly
degenerate dark states along paths restricted by the parametrization
\begin{eqnarray}
\label{parameters}
\omega_0(s) & = & \sin \theta(s) \sin \varphi(s),\nonumber\\
\omega_1(s) & = & \sin \theta(s) \cos\varphi(s),\nonumber\\
\omega_a(s) & = & \cos\theta(s),
\end{eqnarray} 
where the angles $\theta$ and $\varphi$ parametrize a 2-sphere. 
Explicitly, a loop  $\mathcal{C}$ in parameter space starting 
and ending at $(\omega_0,\omega_1,\omega_a) = (0,0,1)$, yields 
the holonomic rotation gate  
\begin{eqnarray} 
u[\mathcal{C}] =  
e^{-\Omega (|0\rangle \langle 1| - |1\rangle \langle 0|)} , 
\label{holonomy}
\end{eqnarray}
$\Omega$ being the solid angle swept by $\mathcal{C}$. 

We assume that the system is influenced by an environment which is
sensitive to whether the system is in the state $a$ or not. This 
may be modeled by adding the Lindbladian
\begin{equation}
\label{lindbladian}
V = |a\rangle \langle a|   
\end{equation}
and its concomitant strength $\Gamma$. 

\subsection{Application of the approximation}
First, we notice that there is an arbitrariness in the choice of
eigenbasis of $H(s)$, which can be formulated as a choice of
gauge. This arbitrariness in the choice of gauge is related to the
arbitrariness in the choice of $U(s)$ in Eq.~(\ref{Pvillkor}). Let
$\{|\chi_{k}(s)\rangle\}_{k}$ be an instantaneous orthonormal
eigenbasis of $H(s)$. Given such a basis one may construct a family $U(s)$ by
\begin{equation}
\label{gaugeochu}
U(s) = U_{0}\sum_{k} |\chi_{k}(0)\rangle\langle\chi_{k}(s)|,
\end{equation}
where $U_{0}$ is a fixed unitary operator such that $[U_{0},P_{n}(0)]
= 0$ for all $n$. Every family $U(s)$ constructed via
Eq.~(\ref{gaugeochu}) is unitary and satisfies
Eq.~(\ref{Pvillkor}). Moreover, every family $U(s)$ that satisfies
Eq.~(\ref{Pvillkor}) can be reached via Eq.~(\ref{gaugeochu}) for some
choice of instantaneous orthonormal eigenbasis
$\{|\chi_{k}(s)\rangle\}_{k}$ and $U_{0}$. As the present
approximation is independent of the choice of allowed $U(s)$, it
follows that the approximation is also
independent of the choice of gauge.

Here we briefly describe a procedure to put the approximate master
equation into matrix form. As the present application only regards
the computational subspace we disregard the off-diagonal terms of the
approximate solution, and only consider Eq.~(\ref{diagonal}). In the
present case there are three diagonal terms, corresponding to the dark
subspace and the two bright states. In order to write
Eq.~(\ref{diagonal}) on matrix form we first choose an  instantaneous
orthonormal eigenbasis $\{|\chi_{k}(s)\rangle\}_{k}$ of $H(s)$, from 
which one can construct $U(s)$ via Eq.~(\ref{gaugeochu}), with
$U_{0} =
\hat{1}$. Define
\begin{equation}
\label{repr}
\boldsymbol{\rho}^{a} = 
\left(\begin{array}{c c c c c c}\rho^{a}_{11}&\rho^{a}_{12}&
\rho^{a}_{21}&\rho^{a}_{22}&\rho^{a}_{33}&\rho^{a}_{44}\end{array}\right)^t, 
\end{equation}
where $\rho_{kl}^{a}(s) =
\langle\chi_{k}(0)|\widetilde{\rho}^{a}(s)|\chi_{l}(0)\rangle$, and
where $\widetilde{\rho}^{a}(s)$ is the solution of
Eq.~(\ref{diagonal}). Note that we here use the initial eigenbasis
$\{|\chi_{k}(0)\rangle\}_{k}$. This is related to the fact that
Eq.~(\ref{diagonal}) is written in the ``rotated frame'', as described
by Eqs.~(\ref{change}) and (\ref{Pvillkor}). When inserting $U(s)$
into Eq.~(\ref{diagonal}) the result can be written as
\begin{equation}
\label{eqn:matrixapprox}
\dot{\boldsymbol{\rho}}^{a} =  
\boldsymbol{M}^{a}(s)\boldsymbol{\rho}^{a}. 
\end{equation}
If the instantaneous eigenbasis is chosen to be
\begin{widetext}
\begin{eqnarray}
\label{enkel}
|\chi_1(s)\rangle & = & \cos \varphi(s) |0\rangle - \sin \varphi(s)
|1\rangle, 
\nonumber \\ 
|\chi_2(s)\rangle & = & \sin \varphi(s) \cos
\theta(s) |0\rangle + \cos \varphi(s) \cos \theta(s) |1\rangle - \sin
\theta(s) |a\rangle, 
\nonumber \\ 
|\chi_3(s)\rangle & = & \frac{1}{\sqrt{2}} \Big( 
\sin \varphi(s) \sin \theta(s) |0\rangle + \cos \varphi(s) \sin
\theta(s) |1\rangle + \cos \theta(s) |a\rangle +
|e\rangle \Big) , 
\nonumber \\ 
|\chi_4(s)\rangle & = & \frac{1}{\sqrt{2}} \Big( \sin
\varphi(s) \sin \theta(s) |0\rangle + \cos \varphi(s) \sin \theta(s)
|1\rangle + \cos \theta(s) |a\rangle - |e\rangle \Big) ,
\end{eqnarray}
then, with the Hamiltonian in Eq.~(\ref{Hamiltonian}) and the
Lindbladian in Eq.~(\ref{lindbladian}), one obtains
\begin{equation}
\label{Mapprox}
\boldsymbol{M}^{a} = \!\left(\!\begin {array}{cccccc}
0 & -\frac{d\varphi}{ds}\cos\theta(s) &
-\frac{d\varphi}{ds}\cos\theta(s) & 0 & 0 & 0\\
\noalign{\medskip}\frac{d\varphi}{ds}\cos\theta(s)  &
 -\frac{\Gamma T}{2}\sin^2\theta(s) & 0 &
 -\frac{d\varphi}{ds}\cos\theta(s) & 0 & 0\\
\noalign{\medskip}\frac{d\varphi}{ds}\cos\theta(s) 
 & 0 & -\frac{\Gamma T}{2}\sin^2\theta(s) &
 -\frac{d\varphi}{ds}\cos\theta(s) & 0 & 0\\ 0 &
 \frac{d\varphi}{ds}\cos\theta(s) &
\frac{d\varphi}{ds}\cos\theta(s) & -f(s) & \frac{1}{2}f(s) &
\frac{1}{2}f(s)\\ 0 & 0 & 0 & \frac{1}{2}f(s) & -\frac{1}{4}g(s) &
\frac{\Gamma T}{4}\cos^{4}\theta(s)\\ 0 & 0 & 0 & \frac{1}{2}f(s) &
\frac{\Gamma T}{4}\cos^{4}\theta(s) & -\frac{1}{4}g(s)
\end {array}\!\right). 
\end{equation}
\end{widetext}
Here, $f(s) = \Gamma T \sin^2\!\theta(s)\cos^{2}\!\theta(s)$ and
$g(s)=\Gamma T \left[1+\sin^2\!\theta(s)\right]\cos^{2}\!\theta(s)$.

From the above analysis it follows that the solutions of 
Eq.~(\ref{diagonal}) can be written $\widetilde{\rho}^{(nn)}(s) =
\sum_{kl}\rho_{kl}^{a}(s)|\chi_{k}(0)\rangle\langle \chi_{l}(0)|$,
where the sum over $k,l$ spans the appropriate elements for each
$n$. Since these operators are written in the ``rotated frame'', it is
appropriate to invert this transformation to more easily analyze the
gate operation. By inverting the transformation in Eq.~(\ref{change})
one finds that these operators can be written 
\begin{eqnarray}
U^\dagger(s)\widetilde{\rho}^{(nn)}(s)U(s) & = & 
\sum_{kl}\rho_{kl}^{a}(s)|\chi_{k}(s)\rangle\langle \chi_{l}(s)|\nonumber\\
& = & P_{n}(s)\rho(s)P_{n}(s),
\end{eqnarray}
where $\rho(s)$ is the solution of
Eq.~(\ref{totalaapp}).  

There are some subtleties associated with the choice of basis and the
usual difficulty with spherical coordinates, viz., that $\varphi$ is
not defined at the north and south pole of parameter space. In fact,
with the choice of basis in Eq.~(\ref{enkel}), the gauge potential
$Z(s)$ has singularities at both poles. Nevertheless, if we avoid
loops around the poles and take appropriate limits if we wish to
approach the poles, then this gauge is unproblematic. Another
possibility is to rotate the dark instantaneous eigenvectors as
\begin{eqnarray} 
|\chi_k\rangle \rightarrow |\chi'_k\rangle & = & 
\sum_{j=1}^2 |\chi_j\rangle \langle \chi_j|W|\chi_k \rangle , \ k=1,2, 
\nonumber \\ 
W & = & e^{\varphi \big( |\chi_1 \rangle \langle \chi_2| - 
|\chi_2 \rangle \langle \chi_1| \big)} . 
\label{nordbra}
\end{eqnarray}
With this basis one obtains a gauge where the vector potential is well
defined except at the south pole of parameter space. 
The rotated dark states does, however, give a system of differential 
equations too extensive to be presented explicitly here. 
 
\begin{figure}[ht]
\includegraphics[width = 6cm]{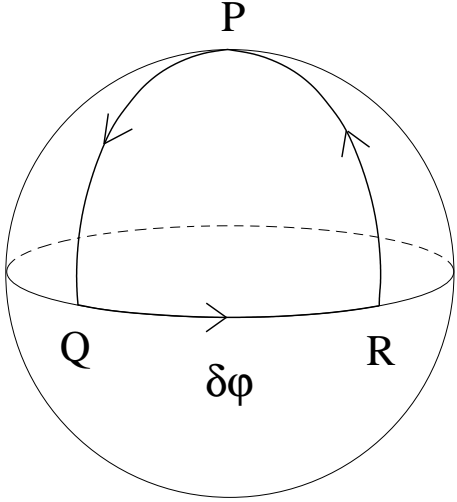}
\caption{\label{fig:path} Path in parameter space starting and 
ending at P. The duration of the path segments P$\rightarrow$Q,
Q$\rightarrow$R, and R$\rightarrow$P is $T_1$, $T_2$, and $T_3$,
respectively, as described in Eq.~(\ref{eqn:orangepath}). 
$\delta \varphi$ is the opening angle in the equatorial plane. 
Note that for this path the enclosed solid angle equals $\delta\varphi$.}
\end{figure}

We restrict the parametrization to 
\begin{equation}
\varphi(s) = a\, s + b, \quad \theta(s) = c\, s + d,
\quad a,b,c,d \in \mathbb{R},\label{eqn:resres}
\end{equation}
and the paths in parameter space to half ``orange slices'' 
\begin{eqnarray}
\label{eqn:orangepath}
(\varphi=0,\theta=0,t=0) & \rightarrow 
& (0,\pi/2,T_1) 
\nonumber \\  
& \rightarrow & (\delta \varphi,\pi/2,T_2+T_1) 
\nonumber \\ 
& \rightarrow &(\delta \varphi,0,T_3+T_2+T_1) 
\nonumber \\ 
& \rightarrow & (0,0,T_4+T_3+T_2+T_1) , 
\nonumber \\ 
\end{eqnarray} 
where $T_1,\ldots,T_4$ are the run-times for the path segments, see
Fig.~\ref{fig:path}. Note that the fourth path originates from the
deformation of a well defined square to the orange slice on the
parameter sphere. In the limit $\theta \rightarrow 0$ the fourth path
is reduced to a single point at the north pole. This implies that
$T_{4}$ can be set to zero without loss of adiabaticity. Note that in
the present decoherence model a nonzero $T_{4}$ only affects the two
bright states.

For the initial state vector $|\Psi\rangle = \cos(x/2)|0\rangle +
e^{-iy}\sin(x/2)|1\rangle$, the output state of the approximation
projected onto the computational space may be written as
\begin{eqnarray}
\label{approximerade}
\boldsymbol{\rho}^a_{\mathrm{out}} & = & 
\boldsymbol{u} [\mathcal{C}] \boldsymbol{\rho}'
\boldsymbol{u}^{\dagger} [\mathcal{C}] , 
\nonumber\\
\boldsymbol{\rho}' & = & \left( \begin {array}{cc} 
\frac{1}{2}+\frac{1}{2}\,\cos x  & f_1 \\
f_1^* & \left(\frac{1}{2}-\frac{1}{2}\cos x \right) f_2
\end {array}\right) ,
\end{eqnarray}
where 
\begin{eqnarray}
f_1 & = & \frac{1}{2} e^{-\frac{1}{4}\Gamma \left( T_3+2T_2+T_1
\right)} e^{-iy}\sin x,\\ f_2 & = & \frac{1}{3} +
\frac{2}{3}e^{-\frac{3}{16}\Gamma(T_3 +T_1)} , 
\end{eqnarray}
and $\boldsymbol{u} [\mathcal{C}]$ is the holonomy
Eq.~(\ref{holonomy}) in the $\{ |0\rangle , |1\rangle \}$ basis. Thus,
the output is determined by the holonomy transformation of the
$\delta\varphi$-independent $\boldsymbol{\rho}'$. It is worth to point
out that this feature is due to the particular choice of parametrization
and the chosen loop in parameter space, and
not some intrinsic property of the decoherence. Furthermore, in
addition to destroying the superpositions between the computational
states the decoherence also gives an intensity loss from the
computational space. This intensity loss arises as the states
corresponding to $\chi_2(s)$, $\chi_3(s)$, and $\chi_4(s)$ decoheres
into mixed states, as all three of them contain the $a$ state.

As a final observation we note that Eq.~(\ref{eqn:matrixapprox}) can
be obtained more or less directly from Eq.~(\ref{basekv}).  We may
represent Eq.~(\ref{basekv}) using an instantaneous orthonormal
eigenbasis $\{|\chi_{k}(s)\rangle\}_{k}$ as
\begin{equation}
\label{totalaekv}
\dot{\boldsymbol{\rho}} = \boldsymbol{M}(s)\boldsymbol{\rho},
\end{equation}
where $\boldsymbol{M}(s)$ is a $16\times 16$ matrix, and where 
\begin{equation}
\boldsymbol{\rho} \equiv \left(\begin{array}{c c c c c c c c}\rho_{11}
&\rho_{12}&\cdots&\rho_{14}&\rho_{21}&
\cdots&\rho_{44}\end{array}\right)^t 
\end{equation}
with $\rho_{kl}(s) = \langle \chi_k (s)|\rho(s)|\chi_l (s) \rangle , \
k,l=1,\ldots,4$.  Due to Eq.~(\ref{change}) it follows that
Eq.~(\ref{totalaekv}) is also obtained if we instead represent
Eq.~(\ref{huvudekv}) using the $\{|\chi_{k}(0)\rangle\}_{k}$ basis,
where we again assume that $U(s)$ is constructed via
Eq.~(\ref{gaugeochu}) with $U_{0}=\hat{1}$. Equation (\ref{nlan}) is
obtained by removing couplings from Eq.~(\ref{huvudekv}). For the
chosen basis, this corresponds to a removal of off-diagonal elements
in $\boldsymbol{M}(s)$, such that the new approximate matrix can be
arranged in a block diagonal form. Each of these diagonal blocks
corresponds to a collection of coupled terms. One block corresponds to
the diagonal terms, and there is one block for each collection of
off-diagonal terms that couples among themselves, as determined by
$g_{klk'l'}$. If one is interested in the evolution of a particular
collection of coupled terms, then the approximate equation is obtained
if one removes those rows and columns from $\boldsymbol{M}(s)$ that
correspond to terms not included in the collection.  In the present
example, the matrix $\boldsymbol{M}^{a}(s)$ in Eq.~(\ref{Mapprox}) is
obtained if we use the basis in Eq.~(\ref{enkel}) to represent the
exact master equation, and remove those rows and columns from
$\boldsymbol{M}(s)$ that correspond to the off-diagonal terms.

\subsection{\label{sec.numeric}Numerical analysis}
We compare the approximate solution in Eq.~(\ref{approximerade}) with
a numerical solution of Eq.~(\ref{totalaekv}) in the Hadamard case
$\Omega = \pi/4$ by putting $\delta \varphi = \pi/4$. For the
calculation we have used the gauge where the vector potential is well
defined at the north pole. We further put $T_{4} = 0$. We distribute
the run-time $T$ proportionally to the length of the three circle
segments, i.e., $T_{1} = T_{3} = 2T/5$, $T_{2} = T/5$.

In the numerical treatment of the evolution we decompose the 
interval $[0,T]$ into subintervals with step size $\Delta t$, on 
which $\boldsymbol{M}(t)$ is taken to be constant. The resulting 
approximate evolution is on the form $\boldsymbol{\rho}_{K} =
\big[ \Pi_{k=0}^{K} \exp \bm{(} \Delta t \boldsymbol{M}(t_{k}) 
\bm{)} \big] \boldsymbol{\rho}_{0}$. The step size $\Delta t = 0.01$ 
is used. The step size $\Delta t = 0.005$ has been tested, without 
any significant change of the result.

For quantum gates, the relevant error is that at the end-point. This 
error may contain a contribution in form of an intensity loss out of
the computational subspace. To detect this intensity loss, we use the 
quantities 
\begin{eqnarray}
I(T) & = & 1-\Tr \big( P\varrho (T) \big) , 
\nonumber \\ 
I^a(T) & = & 1-\Tr \big( P\varrho^a (T) \big)  
\label{intensityloss}
\end{eqnarray} 
with $P$ the projector onto the computational subspace spanned by
$|0\rangle$ and $|1\rangle$. The error within this 
subspace is analyzed in terms of the fidelity \cite{Uhlmann,Jozsa} 
\begin{eqnarray}
 & & D(\varrho_{\mathrm{norm}} (T), \varrho^a_{\mathrm{norm}} (T)) 
\nonumber \\ 
 & & \equiv \Tr \sqrt{\sqrt{\varrho_{\mathrm{norm}} (T)} 
\varrho^a_{\mathrm{norm}} (T)\sqrt{\varrho_{\mathrm{norm}} (T)}} ,  
\label{fidelity}
\end{eqnarray}
where 
\begin{eqnarray} 
\varrho_{\mathrm{norm}} (T) & = & \rho_{\mathrm{norm}} (1)  = 
\frac{P \varrho(T) P}{\Tr \big( P\varrho(T) \big)} , 
\nonumber \\ 
\varrho^a_{\mathrm{norm}} (T) & = & \rho^a_{\mathrm{norm}} (1) = 
\frac{P\varrho^{a}(T)P}{\Tr \big( P\varrho^{a}(T) \big)} ,   
\label{normstates}
\end{eqnarray} 
are the normalized outputs of the exact and the approximate evolution,
respectively. This normalization may correspond to a post-selection 
procedure. 

\begin{figure}[ht]
\includegraphics[width = 8cm]{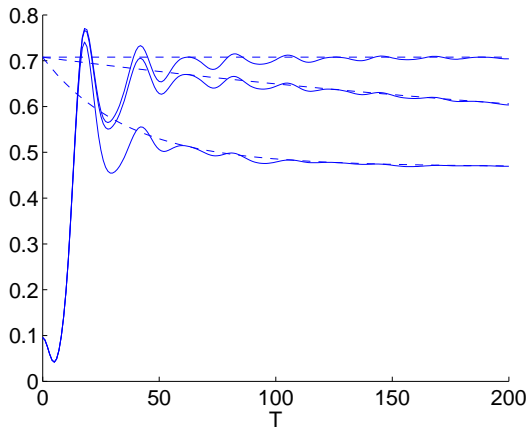}
\caption{\label{fig:element}
The solid lines show the value of the matrix element $\langle
1|\varrho(T)|1\rangle$ of the output density operator of the exact
equation, as a function of $T$. The run-time $T$ is measured in units
of $\omega^{-1}$, defined in Eq.~(\ref{smomedef}).  The dashed lines
show the corresponding value $\langle
1|\varrho^{a}(T)|1\rangle$. Counted from the top and down, the
solid-dashed line pairs correspond to $\Gamma =0$, $\Gamma = 0.01$,
and $\Gamma = 0.1$, respectively.  The initial state is pure, with
polar angle $x = \pi/5$ and azimuthal angle $y =3\pi/4$ on the Bloch
sphere. The horizontal dashed line corresponds to the ordinary
adiabatic approximation for the closed evolution case. One may note
that the rate at which the exact solution approaches the approximate
solution appears to be rather independent of the strength $\Gamma$ of
the decoherence.}
\end{figure}

\begin{figure}[ht]
\includegraphics[width = 8cm]{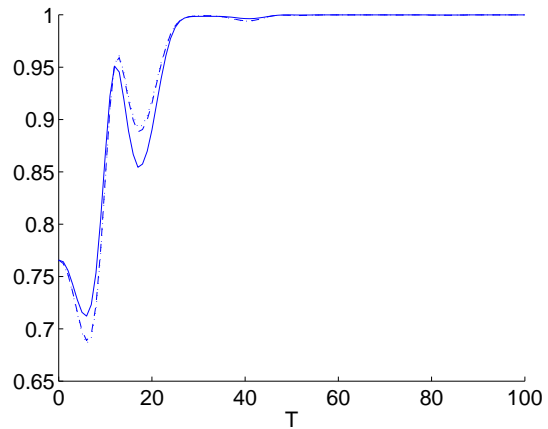}
\caption{\label{fig:fidelity}
These graphs highlight another aspect of the same series of
calculations as in Fig.~\ref{fig:element}. They show the error between
the exact evolution and the approximate evolution, in form of the
normalized fidelity $D$ defined in Eqs.~(\ref{fidelity}) and
(\ref{normstates}), as a function of $T$. The run-time $T$ is given in
units of $\omega^{-1}$, defined in Eq.~(\ref{smomedef}). Note the
different range of the run-time compared to the other plots. Here the
dotted line corresponds to $\Gamma = 0$, the dashed line to $\Gamma =
0.01$ and the solid line to $\Gamma = 0.1$. Note that the dotted and
the dashed lines almost coincide. These graphs indicate that the
distance between the approximate and exact evolution, within the
computational subspace, decreases with the run-time $T$ at a rate more
or less independent of $\Gamma$.}
\end{figure}

\begin{figure}
\includegraphics[width = 8cm]{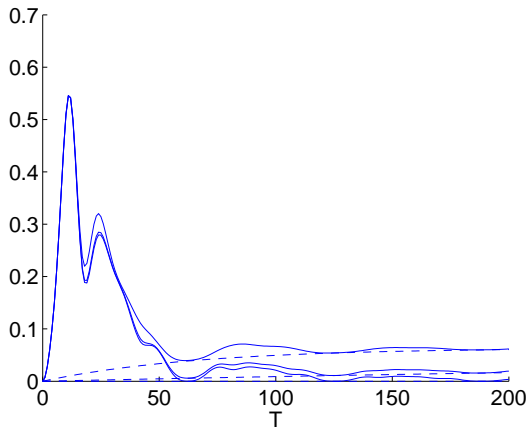}
\caption{\label{fig:loss}
The intensity losses, as defined in Eq.~(\ref{intensityloss}), out of
the computational subspace for the same series of calculations as in
Figs.~\ref{fig:element} and
\ref{fig:fidelity}. 
The solid lines show the intensity losses $I$ of the exact evolution,
and the dashed the intensity losses $I^{a}$ of the approximate
evolution, as a function of $T$. The run-time $T$ is given in units of
$\omega^{-1}$, defined in Eq.~(\ref{smomedef}). Here the lowermost
pair of curves correspond to $\Gamma = 0$. Note that for $\Gamma = 0$
the loss is identically zero for the approximate evolution. The
uppermost pair of curves corresponds to $\Gamma = 0.1$, and the pair
in the middle corresponds to $\Gamma = 0.01$.}
\end{figure}

In Fig.~\ref{fig:element}, we show $\langle 1|\varrho(T)|1\rangle$ and
$\langle 1|\varrho^{a}(T)|1\rangle$, for $\Gamma = 0,0.01,0.1$. 
We have chosen the initial state vector $|\Psi\rangle = \cos(x/2)|0\rangle
+ e^{-iy}\sin(x/2)|1\rangle$ with $x = \pi/5$ and $y = 3\pi/4$. The
corresponding normalized fidelity $D$ is shown in
Fig.~\ref{fig:fidelity} and the intensity losses $I(T)$ and $I_a(T)$
are shown in Fig.~\ref{fig:loss}. These simulations indicate that for
this model system the error seems to decrease with increasing run-times
$T$ at a rate more or less equal to the ordinary adiabatic
approximation in the closed case, independent of the strength $\Gamma$
of the decoherence process. We have confirmed this finding for other
input states.

\section{\label{sec:range}Range of applicability}
\begin{figure}
\includegraphics[width = 8cm]{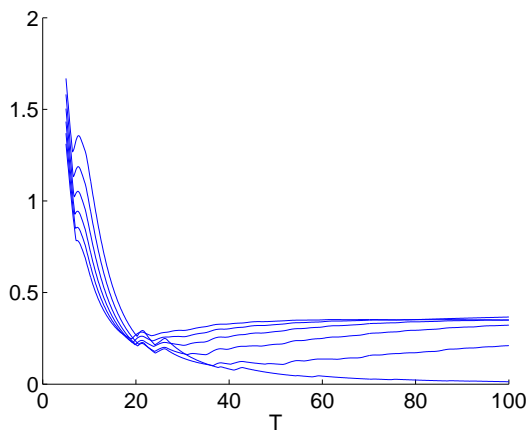}
\caption{\label{fig:exemp} The maximum error in the Hilbert-Schmidt 
norm $\max_{s\in[0,1]}||\rho(s)-\rho^{a}(s)||$ between the solution
$\rho(s)$ of Eq.~(\ref{dekoh}) and the solution $\rho^{a}(s)$ of the
approximate equation as a function of the run-time $T$, the latter
measured in arbitrary units. The plots are generated for one random
instance of $H_{0}$, $Z$, $A$, and initial state, for a four
dimensional Hilbert space. Each curve corresponds to a value of
$\Gamma$, and shows the maximum error as a function of $T$. To the
right of the figure the curves correspond to, counted from the bottom
and up, to $\Gamma = 0, 0.002, 0.004, 0.006, 0.008, 0.01$. As seen,
the error for the adiabatic approximation in the closed case ($\Gamma
=0$) seems to tend to zero as $T$ increases, while the other cases
appear to have a minimum error for a certain $T$.}
\end{figure}

The analysis in Sec.~\ref{sec:weak}
suggests that, for a given $\Gamma$, the error bound in Eq.~(\ref{cond}) 
has a minimum for some value of the run-time $T$.  
It is quite straightforward to obtain
an example of a system where this appears to be the case. One may
consider a time-dependent Hamiltonian of the form
\begin{equation}
H(s) = e^{-isZ}H_{0}e^{isZ},
\end{equation}
where $H_{0}$ and $Z$ are fixed Hermitian operators. The
spectrum of this Hamiltonian is fixed, but the eigenbasis rotates. One
may consider the master equation
\begin{equation}
\label{dekoh}
\dot{\rho} = -iT[H(s),\rho] -\Gamma T [A,[A,\rho]],
\end{equation}
where $A$ is a fixed Hermitian operator. The double commutator in the
above equation causes decoherence with respect to the eigenbasis of
$A$. We have chosen a four dimensional Hilbert space and have
generated $H_{0}$, $Z$, and $A$, as well as the pure initial state,
randomly. Figure \ref{fig:exemp} shows the maximum error in the
Hilbert-Schmidt norm $\max_{s\in[0,1]}||\rho(s)-\rho^{a}(s)||$ for
various choices of $T$ and $\Gamma$.  As seen in Fig.~\ref{fig:exemp}
we indeed seem to have the expected behavior of the approximation. In
the ideal case, $\Gamma = 0$, the error appears to go to zero as $T$
increases, while for non-vanishing $\Gamma$ there seems to be a
minimum error.

However, the error does not always seem to behave in this manner.  In
the example presented in Sec.~\ref{sec.numeric} there is no trace of
this minimum error. Rather the error seems to vanish for large $T$ for
any value of $\Gamma$. In other words, the error bounds derived in
Sec.~\ref{sec:weak} appears to be unnecessarily pessimistic in some
cases. We here put forward some reasons why this may be the case.

One aspect is the question of which error to consider. In
Sec.~\ref{sec:weak} we considered the maximum deviation
between the exact and approximate solution during the whole evolution,
while in Sec.~\ref{sec.numeric}, the relevant error was taken at the
end of the evolution. In some cases the maximum deviation need not
occur at the end of the evolution, which may cause the ``end point
error'' to be smaller than the maximum deviation. One may also note
that Sec.~\ref{sec.numeric} focused on one single diagonal term
of the total density operator and that the error for this part may
be smaller than the total error.

Another reason for the approximation to be accurate under
wider conditions is if the dissipator $D_{s}$ is such that it does not
couple off-diagonal terms to diagonal terms, or off-diagonal terms to
other off-diagonal terms. Under such conditions the dissipator is
unaffected by the approximation and it seems reasonable that the
approximation should have a wider range of applicability. 
However, this cannot be the sole reason, as is indicated by the
results in Sec.~\ref{sec.numeric}, since the dissipator used
(i.e., decoherence with pointer state $a$) does belong to the class
of dissipators that do couple the diagonal and off-diagonal terms.
Suppose, however, that the evolution is such that the magnitude of the
off-diagonal terms tends to decrease with increasing run-times. For
example, this may occur if a decoherence or relaxation process acts
suitably in relation to the instantaneous eigenspaces. Consider the
diagonal terms: even if there would be a coupling to the off-diagonal
terms, the importance of this coupling naturally diminishes if the
off-diagonal terms tends to decrease in magnitude. If the open system
process is such that it tends to suppress the off-diagonal terms, it
thus seems reasonable to expect that the approximate equation should
be accurate at large run-times. This reduction of off-diagonal terms
reasonably should be more relevant for the end point error than for
the maximum deviation.  Another reason for the end-point error
to vanish is if the approximate and exact equations have a common
asymptotic state.

There is clearly room for further investigations of when and why the
present approximation is accurate beyond the joint limit of slow
change and weak open system effects.

\section{\label{sec:sum}Conclusions}

We present an adiabatic approximation scheme for weakly open systems.
Contrary to the adiabatic approximation for closed systems, the
presence of open system effects introduces a coupling between the
instantaneous eigenspaces of the time-dependent, possibly degenerate,
Hamiltonian.  We show that the present approximation can be obtained
as a slow-change weak open-system limit, in the sense that the
 time scale inversely proportional to
the strength of the open system effect puts an upper limit on
the run-time. In the ideal case of closed systems this limiting
time scale becomes infinite, and the ordinary adiabatic approximation
\cite{Messiah} is retained.

We demonstrate the approximation scheme for a non-Abelian holonomic
implementation of a Hadamard gate, exposed to a decoherence
process. We compare the approximation with numerically obtained
solutions of the exact master equation. 
These calculations indicate that the error between the approximate and
the exact evolution decreases with increasing run-time at a rate more
or less independent of the strength of the decoherence process.
This result suggests that the approximation scheme may have a wider
range of applicability than the weak open-system limit.


\begin{thebibliography}{99}
\expandafter\ifx\csname natexlab\endcsname\relax\def\natexlab#1{#1}\fi
\expandafter\ifx\csname bibnamefont\endcsname\relax
  \def\bibnamefont#1{#1}\fi
\expandafter\ifx\csname bibfnamefont\endcsname\relax
  \def\bibfnamefont#1{#1}\fi
\expandafter\ifx\csname citenamefont\endcsname\relax
  \def\citenamefont#1{#1}\fi
\expandafter\ifx\csname url\endcsname\relax
  \def\url#1{\texttt{#1}}\fi
\expandafter\ifx\csname urlprefix\endcsname\relax\def\urlprefix{URL }\fi
\providecommand{\bibinfo}[2]{#2}
\providecommand{\eprint}[2][]{\url{#2}}

\bibitem[{\citenamefont{Messiah}(1962)}]{Messiah}
\bibinfo{author}{\bibfnamefont{A.}~\bibnamefont{Messiah}},
  \emph{\bibinfo{title}{Quantum Mechanics}} (\bibinfo{publisher}{North-Holland,
  Amsterdam}, \bibinfo{year}{1962}), Vol. 2.

\bibitem[{\citenamefont{Zanardi and Rasetti}(1999)}]{Zanardi}
\bibinfo{author}{\bibfnamefont{P.}~\bibnamefont{Zanardi}} \bibnamefont{and}
  \bibinfo{author}{\bibfnamefont{M.}~\bibnamefont{Rasetti}},
  \bibinfo{journal}{Phys. Lett. A} \textbf{\bibinfo{volume}{264}},
  \bibinfo{pages}{94} (\bibinfo{year}{1999}).

\bibitem[{\citenamefont{Farhi et~al.}(2000)\citenamefont{Farhi, Goldstone,
  Gutmann, and Sipser}}]{FarhiGold}
\bibinfo{author}{\bibfnamefont{E.}~\bibnamefont{Farhi}},
  \bibinfo{author}{\bibfnamefont{J.}~\bibnamefont{Goldstone}},
  \bibinfo{author}{\bibfnamefont{S.}~\bibnamefont{Gutmann}}, \bibnamefont{and}
  \bibinfo{author}{\bibfnamefont{M.}~\bibnamefont{Sipser}},
  \eprint{quant-ph/0001106}.

\bibitem[{\citenamefont{Farhi et~al.}(2001)\citenamefont{Farhi, Goldstone,
  Gutmann, Lapan, Lundgren, and Preda}}]{Farhi}
\bibinfo{author}{\bibfnamefont{E.}~\bibnamefont{Farhi}},
  \bibinfo{author}{\bibfnamefont{J.}~\bibnamefont{Goldstone}},
  \bibinfo{author}{\bibfnamefont{S.}~\bibnamefont{Gutmann}},
  \bibinfo{author}{\bibfnamefont{J.}~\bibnamefont{Lapan}},
  \bibinfo{author}{\bibfnamefont{A.}~\bibnamefont{Lundgren}}, \bibnamefont{and}
  \bibinfo{author}{\bibfnamefont{D.}~\bibnamefont{Preda}},
  \bibinfo{journal}{Science} \textbf{\bibinfo{volume}{292}},
  \bibinfo{pages}{472} (\bibinfo{year}{2001}).

\bibitem[{\citenamefont{Romero et~al.}(2002)\citenamefont{Romero, Pinto, and
  Thomaz}}]{rompinto}
\bibinfo{author}{\bibfnamefont{K.~M.~F.} \bibnamefont{Romero}},
  \bibinfo{author}{\bibfnamefont{A.~C.~A.} \bibnamefont{Pinto}},
  \bibnamefont{and} \bibinfo{author}{\bibfnamefont{M.~T.}
  \bibnamefont{Thomaz}}, \bibinfo{journal}{Physica A}
  \textbf{\bibinfo{volume}{307}}, \bibinfo{pages}{142} (\bibinfo{year}{2002}).

\bibitem[{\citenamefont{Pinto et~al.}(2002)\citenamefont{Pinto, Romero, and
  Thomaz}}]{pintorom}
\bibinfo{author}{\bibfnamefont{A.~C.~A.} \bibnamefont{Pinto}},
  \bibinfo{author}{\bibfnamefont{K.~M.~F.} \bibnamefont{Romero}},
  \bibnamefont{and} \bibinfo{author}{\bibfnamefont{M.~T.}
  \bibnamefont{Thomaz}}, \bibinfo{journal}{Physica A}
  \textbf{\bibinfo{volume}{311}}, \bibinfo{pages}{169 } (\bibinfo{year}{2002}).

\bibitem[{\citenamefont{Wilczek and Zee}(1984)}]{Wilzee}
\bibinfo{author}{\bibfnamefont{F.}~\bibnamefont{Wilczek}} \bibnamefont{and}
  \bibinfo{author}{\bibfnamefont{A.}~\bibnamefont{Zee}},
  \bibinfo{journal}{Phys. Rev. Lett} \textbf{\bibinfo{volume}{52}},
  \bibinfo{pages}{2111} (\bibinfo{year}{1984}).

\bibitem[{\citenamefont{Sarandy and Lidar}(2005{\natexlab{a}})}]{sarlidar}
\bibinfo{author}{\bibfnamefont{M.~S.} \bibnamefont{Sarandy}} \bibnamefont{and}
  \bibinfo{author}{\bibfnamefont{D.~A.} \bibnamefont{Lidar}},
  \bibinfo{journal}{Phys. Rev. A} \textbf{\bibinfo{volume}{71}},
  \bibinfo{pages}{012331} (\bibinfo{year}{2005}{\natexlab{a}}).

\bibitem[{\citenamefont{Sarandy and Lidar}(2005)\citenamefont{Sarandy and Lidar}}]{sali}
\bibinfo{author}{\bibfnamefont{M.~S.}~\bibnamefont{Sarandy}} \bibnamefont{and}
  \bibinfo{author}{\bibfnamefont{D.~A.}~\bibnamefont{Lidar}},
  \eprint{quant-ph/0502014}.

\bibitem{Aab}
\bibinfo{author}{\bibfnamefont{J.}~\bibnamefont{{\AA}berg}},
  \bibinfo{author}{\bibfnamefont{D.}~\bibnamefont{Kult}}, \bibnamefont{and}
  \bibinfo{author}{\bibfnamefont{E.}~\bibnamefont{Sj\"oqvist}},
  \bibinfo{journal}{Phys. Rev. A} \textbf{\bibinfo{volume}{71}},
  \bibinfo{pages}{060312(R)} (\bibinfo{year}{2005}{\natexlab{a}}).

\bibitem{remark1} By putting $\Gamma =0$ in Eq.~(\ref{totalaapp})
we obtain the Schr\"odinger equation $|\dot{\psi}\rangle =
-i[TH(s)+Q(s)]|\psi\rangle$. If $H(s)$ is
nondegenerate with orthonormal eigenbasis
$\{|\chi_{k}(s)\rangle\}_{k}$, then we may write $|\psi(s)\rangle =
\sum_{k}c_{k}(s)|\chi_{k}(s)\rangle$. The resulting equations
for the coefficients $c_{k}$ become $\dot{c}_{k}(s) = [-iTE_{k}(s)
+\langle\dot{\chi}_{k}(s)|\chi_{k}(s)\rangle]c_{k}(s)$.

\bibitem[{\citenamefont{Marsden and Hoffman}(1987)}]{complex}
\bibinfo{author}{\bibfnamefont{J.~E.} \bibnamefont{Marsden}} \bibnamefont{and}
  \bibinfo{author}{\bibfnamefont{M.~J.} \bibnamefont{Hoffman}},
  \emph{\bibinfo{title}{Basic Complex Analysis}} (\bibinfo{publisher}
  {Freeman, New York}, \bibinfo{year}{1987}).

\bibitem[{\citenamefont{Churchill}(1963)}]{RimLeb}
\bibinfo{author}{\bibfnamefont{R.~V.} \bibnamefont{Churchill}},
  \emph{\bibinfo{title}{Fourier Series and Boundary Value Problems}}
  (\bibinfo{publisher}{McGraw-Hill, New York}, \bibinfo{year}{1963}).

\bibitem[{\citenamefont{Rellich}(1969)}]{Rell}
\bibinfo{author}{\bibfnamefont{F.}~\bibnamefont{Rellich}},
  \emph{\bibinfo{title}{Perturbation Theory of Eigenvalue Problems}}
  (\bibinfo{publisher}{Gordon and Breach, New York}, \bibinfo{year}{1969}).

\bibitem[{\citenamefont{Amann}(1990)}]{Amann}
\bibinfo{author}{\bibfnamefont{H.}~\bibnamefont{Amann}},
  \emph{\bibinfo{title}{de Gruyter Studies in Mathematics. Ordinary
  Differential Equations}}
  (\bibinfo{publisher}{Walter de Gruyter, Berlin}, \bibinfo{year}{1990}), 
  Vol. 13.

\bibitem{reed} M. Reed and B. Simon, 
\emph{Methods of Modern Mathematical Physics IV: Analysis of Operators} 
(Academic Press, New York, 1978). 

\bibitem{remark2} Note that a similar breakdown of adiabaticity after a
 finite time has previously been reported in Ref.~\cite{sarlidar} and has been
 further discussed in Ref.~\cite{sali}.

\bibitem[{\citenamefont{Lindblad}(1976)}]{Lindblad}
\bibinfo{author}{\bibfnamefont{G.}~\bibnamefont{Lindblad}},
  \bibinfo{journal}{Comm. Math. Phys.} \textbf{\bibinfo{volume}{48}},
  \bibinfo{pages}{119} (\bibinfo{year}{1976}).

\bibitem[{\citenamefont{Gorini et~al.}(1976)\citenamefont{Gorini, Kossakowski,
  and Sudarshan}}]{Vitt}
\bibinfo{author}{\bibfnamefont{V.}~\bibnamefont{Gorini}},
  \bibinfo{author}{\bibfnamefont{A.}~\bibnamefont{Kossakowski}},
  \bibnamefont{and} \bibinfo{author}{\bibfnamefont{E.~C.~G.}
  \bibnamefont{Sudarshan}}, \bibinfo{journal}{J. Math. Phys.}
  \textbf{\bibinfo{volume}{17}}, \bibinfo{pages}{821} (\bibinfo{year}{1976}).

\bibitem[{\citenamefont{Kraus}(1983)}]{Kraus}
\bibinfo{author}{\bibfnamefont{K.}~\bibnamefont{Kraus}},
  \emph{\bibinfo{title}{ States, Effects, and
  Operations}}, Lecture Notes in Physics Vol. 190 (\bibinfo{publisher}{Springer, Berlin},
  \bibinfo{year}{1983}).

\bibitem[{\citenamefont{Alicki and Lendi}(1987)}]{Alle}
\bibinfo{author}{\bibfnamefont{R.}~\bibnamefont{Alicki}} \bibnamefont{and}
  \bibinfo{author}{\bibfnamefont{K.}~\bibnamefont{Lendi}},
  \emph{\bibinfo{title}{Quantum Dynamical Semigroups
  and Applications}}, Lecture Notes in Physics Vol. 286
  (\bibinfo{publisher}{Springer, Berlin}, \bibinfo{year}{1987}).

\bibitem[{\citenamefont{Lendi}(1986)}]{Lendi}
\bibinfo{author}{\bibfnamefont{K.}~\bibnamefont{Lendi}},
  \bibinfo{journal}{Phys. Rev. A} \textbf{\bibinfo{volume}{33}},
  \bibinfo{pages}{3358} (\bibinfo{year}{1986}).


\bibitem[{\citenamefont{Wilkie}(2000)}]{Wilkie}
\bibinfo{author}{\bibfnamefont{J.}~\bibnamefont{Wilkie}},
  \bibinfo{journal}{Phys. Rev. E} \textbf{\bibinfo{volume}{62}},
  \bibinfo{pages}{8808} (\bibinfo{year}{2000}).


\bibitem[{\citenamefont{Breuer}(2004)}]{Breuer}
\bibinfo{author}{\bibfnamefont{H.-P.}~\bibnamefont{Breuer}},
  \bibinfo{journal}{Phys. Rev. A} \textbf{\bibinfo{volume}{70}},
  \bibinfo{pages}{012106} (\bibinfo{year}{2004}).


\bibitem[{\citenamefont{Unanyan et~al.}(1999)\citenamefont{Unanyan, Shore, and
  Bergmann}}]{Unan}
\bibinfo{author}{\bibfnamefont{R.~G.} \bibnamefont{Unanyan}},
  \bibinfo{author}{\bibfnamefont{B.~W.} \bibnamefont{Shore}}, \bibnamefont{and}
  \bibinfo{author}{\bibfnamefont{K.}~\bibnamefont{Bergmann}},
  \bibinfo{journal}{Phys. Rev. A} \textbf{\bibinfo{volume}{59}},
  \bibinfo{pages}{2910} (\bibinfo{year}{1999}).

\bibitem[{\citenamefont{Duan et~al.}(2001)\citenamefont{Duan, Cirac, and
  Zoller}}]{Duan}
\bibinfo{author}{\bibfnamefont{L.~M.} \bibnamefont{Duan}},
  \bibinfo{author}{\bibfnamefont{J.~I.} \bibnamefont{Cirac}}, \bibnamefont{and}
  \bibinfo{author}{\bibfnamefont{P.}~\bibnamefont{Zoller}},
  \bibinfo{journal}{Science} \textbf{\bibinfo{volume}{292}},
  \bibinfo{pages}{1695} (\bibinfo{year}{2001}).

\bibitem[{\citenamefont{Faoro et~al.}(2003)\citenamefont{Faoro, Siewert, and
  Fazio}}]{Faoro}
\bibinfo{author}{\bibfnamefont{L.}~\bibnamefont{Faoro}},
  \bibinfo{author}{\bibfnamefont{J.}~\bibnamefont{Siewert}}, \bibnamefont{and}
  \bibinfo{author}{\bibfnamefont{R.}~\bibnamefont{Fazio}},
  \bibinfo{journal}{Phys. Rev. Lett.} \textbf{\bibinfo{volume}{90}},
  \bibinfo{pages}{028301} (\bibinfo{year}{2003}).

\bibitem[{\citenamefont{Solinas et~al.}(2003)\citenamefont{Solinas, Zanardi,
  Zangh\`\i, and Rossi}}]{solzanzanros}
\bibinfo{author}{\bibfnamefont{P.}~\bibnamefont{Solinas}},
  \bibinfo{author}{\bibfnamefont{P.}~\bibnamefont{Zanardi}},
  \bibinfo{author}{\bibfnamefont{N.}~\bibnamefont{Zangh\`\i}},
  \bibnamefont{and} \bibinfo{author}{\bibfnamefont{F.}~\bibnamefont{Rossi}},
  \bibinfo{journal}{Phys. Rev. A} \textbf{\bibinfo{volume}{67}},
  \bibinfo{pages}{062315} (\bibinfo{year}{2003}).

\bibitem[{\citenamefont{Recati et~al.}(2002)\citenamefont{Recati, Calarco,
  Zanardi, Cirac, and Zoller}}]{RecCal}
\bibinfo{author}{\bibfnamefont{A.}~\bibnamefont{Recati}},
  \bibinfo{author}{\bibfnamefont{T.}~\bibnamefont{Calarco}},
  \bibinfo{author}{\bibfnamefont{P.}~\bibnamefont{Zanardi}},
  \bibinfo{author}{\bibfnamefont{J.~I.} \bibnamefont{Cirac}}, \bibnamefont{and}
  \bibinfo{author}{\bibfnamefont{P.}~\bibnamefont{Zoller}},
  \bibinfo{journal}{Phys. Rev. A} \textbf{\bibinfo{volume}{66}},
  \bibinfo{pages}{032309} (\bibinfo{year}{2002}).

\bibitem[{\citenamefont{Uhlmann}(1976)}]{Uhlmann}
\bibinfo{author}{\bibfnamefont{A.}~\bibnamefont{Uhlmann}},
  \bibinfo{journal}{Rep. Math. Phys} \textbf{\bibinfo{volume}{9}},
  \bibinfo{pages}{273} (\bibinfo{year}{1976}).

\bibitem[{\citenamefont{Jozsa}(1994)}]{Jozsa}
\bibinfo{author}{\bibfnamefont{R.}~\bibnamefont{Jozsa}}, \bibinfo{journal}{J.
  Mod. Opt.} \textbf{\bibinfo{volume}{41}}, \bibinfo{pages}{2315}
  (\bibinfo{year}{1994}).

\end{thebibliography}
\end{document}